\documentclass[manuscript,screen,review]{acmart}

\AtBeginDocument{%
  }

\setcopyright{acmlicensed}
\copyrightyear{2024}
\acmYear{2024}
\acmDOI{XXXXXXX.XXXXXXX}

\acmConference[Conference acronym 'XX]{Make sure to enter the correct
  conference title from your rights confirmation emai}{June 03--05,
  2024}{Woodstock, NY}
\acmISBN{978-1-4503-XXXX-X/18/06}

\usepackage{nicefrac} 
\usepackage{float}
\usepackage{etoolbox}
\apptocmd{\maketitle}{\thispagestyle{empty}}{}{}

\begin{document}

\pagestyle{empty} 
\title{A Comprehensive Survey on Root Cause Analysis in (Micro) Services: Methodologies, Challenges, and Trends}

\pagestyle{empty} 
\author{Tingting Wang}
\email{tingtingwang@seu.edu.cn}
\orcid{0009-0002-8384-1982}
\author{Guilin Qi}
\email{gqi@seu.edu.cn}
\orcid{0000-0003-0150-7236}
\authornote{Corresponding author.}
\affiliation{%
  \institution{Southeast University}
  \city{Nanjing}
  \state{Jiangsu}
  \country{China}
}

\renewcommand{\shortauthors}{Tingting Wang, Guilin Qi  et al.}

\begin{abstract}
\pagestyle{empty} 
The complex dependencies and propagative faults inherent in microservices, characterized by a dense network of interconnected services, pose significant challenges in identifying the underlying causes of issues. Prompt identification and resolution of disruptive problems are crucial to ensure rapid recovery and maintain system stability. Numerous methodologies have emerged to address this challenge, primarily focusing on diagnosing failures through symptomatic data. This survey aims to provide a comprehensive, structured review of root cause analysis (RCA) techniques within microservices, exploring methodologies that include metrics, traces, logs, and multi-model data. It delves deeper into the methodologies, challenges, and future trends within microservices architectures. Positioned at the forefront of AI and automation advancements, it offers guidance for future research directions.
\end{abstract}



\received{2 July 2024}
\received[revised]{xxx xxx xxx}
\received[accepted]{xxx xxx xxx}

\maketitle

\section{Introduction}
\pagestyle{empty} 
The evolution of IT operations has undergone three significant phases, manual operations, DevOps, and AIOps. Initially, IT operations were predominantly manual, relying heavily on human intervention for system monitoring, troubleshooting, and problem resolution. However, with the escalating scale and complexity of systems, the efficacy and precision of manual operations have been increasingly challenged. Subsequently, DevOps was introduced, building upon manual operations and fostering a synergistic collaboration between development and operations. Through automated deployment and continuous integration, DevOps has the capability to expedite the release of new features and rectify issues with greater speed and reliability. Nonetheless, DevOps still necessitates manual involvement in certain complex decision-making processes and tasks. To further mitigate this challenge and enhance cost-effectiveness and efficiency, AIOps leverages machine learning and data analysis to automatically collect and scrutinize vast amounts of IT operation data, enabling real-time monitoring, anomaly detection, fault localization, and automated processing of IT systems. AIOps not only augments the efficiency and accuracy of IT operations but also equips IT operations with the capacity to adapt more effectively to complex and dynamic IT environments, utilizing artificial intelligence and big data technologies.

\setlength{\tabcolsep}{3pt}
\begin{table}
  \caption{Recent Notable Cloud Service and Online Platform Outage Events.[Public domain], via publicly websites. }
  \label{tab:Outage}
  \begin{tabular}{cp{3.3cm}cp{4.7cm}p{4.0cm}}  
   \toprule
Date & Event Name & Duration & Transaction Impact & Cause \\  
\hline  
2024/5/23 & Bing search engine, Microsoft Copilot error  \cite{Microsoft2024} & 5 hours & 429 HTTP error code & Bing’s API error\\  
2024/5/2 & Google Cloud UniSuper disruption \cite{GoogleCloud2024} & 7 days & UniSuper 50 thousand users access 125 billion dollars & Configuration faults\\  
2024/4/8 & Tencent Cloud Console exception \cite{TencentCloud2024} & 87 minutes & Console login exception & Incompatible versions \\  
2023/11/27 & Didi Platform Service interruption \cite{Didi2023} & 12 hours & Ten million orders, 400 million RMB in transactions & Kubernetes cluster and infrastructure issues \\  
2023/11/22 & Google Drive files disappearance \cite{GoogleDrive2023} & six months  & Any uploaded data from May 2023 was gone and Drive had reverted to a state from May. & - \\  
2023/11/12 & Aliyun Global Outage \cite{Aliyun20231127} & 185.76 minutes & Services products, OSS, OTS, SLS, MNS, Cloud Console and management APIs were affected. & Access key (AK) component whitelist contamination \\  
2023/11/8 &  ChatGPT Service Outage  \cite{Mashable2023} & 1.5 hours & ChatGPT of OpenAI and API service interrupted & System overloaded due to far more usage than expected. \\  
2023/6/8 & Outlook Access and Service Issues  \cite{Microsoft20230605} & 12 hours & Users unable to send, receive, or search emails through Outlook.com & Server configuration changes \\  
2023/3/29 & Vipshop Nansha Data Center Failure  \cite{Vipshop2023} & 12 hours & Losses exceeding 100 million RMB, affecting over 80 million customers & Failure of the Nansha IDC cooling system \\  
2023/3/29 & Tencent Wechat, QQ  Service Outage  \cite{Miit2023notice}  & 6 hours & Criticism from the Ministry of Industry and Information Technology & Cooling system anomaly \\  
2023/3/7 &  Twitter Service Outage  \cite{Twitter2023}& 2 hours& Unable to access TweetDeck and 8,000 Twitter outage reports&  API change leads to massive ramifications \\  
2022/12/18 & Aliyun Zone C of the China (Hong Kong) Region Service Outage \cite{Aliyun20221218} & 24 hours & Caused significant public opinion impact & Data center cooling failure triggered the fire sprinkler system \\  
2021/12/7 & Amazon AWS Service Outage \cite{Amazon2021} & 6 hours & Impacted APIs and cloud computing management console functionalities in the US East (N. Virginia) region, affecting Seller Central business &  An automated process to scale capacities resulting in a large surge in connection activity and network congestion. \\  
2021/7/13 & Bilibili Site-wide Anomalies  \cite{Bilibili2021}& 3 hours & Numerous reports of 404 and 502 error codes & Load balancer code defect, rollback operation failed to restore, loss mitigated by rebuilding SLB \\
  \bottomrule
\end{tabular}
\end{table}
As applications migrate to the cloud, containerization, and microservices, the dependencies and invocation relationships between services, as well as between services and components or infrastructure, become increasingly intricate. From a holistic system perspective, the service calls within the system form a network-like structure. When examining an individual service from a vertical perspective, dependencies are organized in a hierarchical manner from top to bottom. Fluctuations in a single service component or service can potentially trigger cascading effects throughout the entire network structure of the system. Although businesses have established a series of defensive measures for stability, including contingency plans, degradation strategies, and disaster recovery, rapid and accurate fault localization at the onset of an issue remains a critical step for subsequent loss prevention and swift system recovery. Inaccurate fault localization can lead to ineffective loss prevention measures, prolonged recovery times, and a negative impact on the service level agreement (SLA) for availability. In practice, significant faults occur frequently, Table ~\ref{tab:Outage} displays the faults recently perceived by users. According to the OpenAI status dashboard, there were 11 incidents in 2024, 14 in February, 34 in March, and 17 and 24 in April and May, respectively; as of now, the total number of reported incidents in 2024 has exceeded 112 \cite{OpenAI2024}. Faults that are perceived by users often lead to significant business losses, increased operational expenses, substantial compensation claims, commercial crises, and heightened risks associated with security and compliance, among a host of other detrimental effects.

\begin{figure}
  \centering
  \includegraphics[width=0.9 \textwidth]{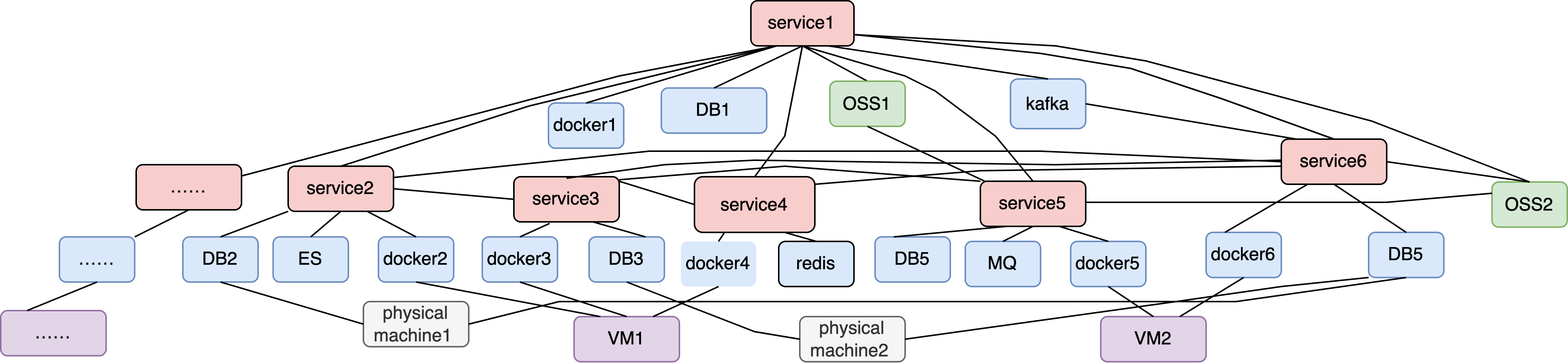}
  \caption{Micro-service-framework of service X}
  \label{fig:microservice-framework}
  \Description{microservice framework of service X, showing its service dependencies, components and infrastructure. The framework illustrates the interconnection between various microservices, highlighting the dependency relationships among them.}
\end{figure}

During a malfunction, faults demonstrate characteristics: destructiveness, dependencies complexity, propagative and recurring, particularly within microservice environment. Faults' destructiveness can result in partial or complete system collapse, severely impacting business operations, disrupting service level agreement and potentially leading to substantial financial and reputational damage. When system availability is compromised, it not only disrupts the continuity of service but also violates service level agreements, which can erode customer trust and satisfaction. The aftermath of such destructive faults often extends beyond immediate technical difficulties, potentially incurring significant financial losses and long-term damage to the company's reputation. Ensuring robust fault tolerance and rapid response mechanisms is therefore crucial to mitigate the destructive potential of system faults. Faults can be diverse and of dependencies complexity, encompassing various facets such as system hardware, components, and network configurations. The complexity of dependencies within microservices architectures (shown in Figure ~\ref{fig:microservice-framework}) significantly increases due to the intricate web of interrelations among components and  services. This "Dependencies Complexity" spans a wide array of elements, including system hardware, individual components, and network configurations. Such multifaceted and intertwined dependencies make the task of diagnosing and resolving faults particularly challenging. Propagative. Because of the dependency’s complexity, faults are propagative \cite{gertler2017fault}. They can propagate from original causes components to another and may even enlarge explosion radius if this affected component is a platform-type like component, which of plenty of call-dependent parties. Thus, the propagation impacts other components of a system and triggers a cascade fault or even anomaly storm \cite{zhao2020understanding}, magnifying the repercussions of a single fault. Recurring. Faults frequently occur in an unpredictable fashion at any given moment and location, posing a challenge for anticipation. Among the frequently failures, about 74.38\% failures are recurring in the investigated applications \cite{li2022actionable}. Because the repetitive and regular occurrence of faults provides historical data guidance for timely diagnosis when faults reoccur.

Considering these characteristics, it is critically important for IT operations teams to rapidly and accurately identify and rectify faults. Root cause analysis represents a pivotal methodology employed within the domain of Artificial Intelligence for IT Operations. This approach facilitates the automated identification of the fundamental causes precipitating a given issue or problem. RCA is achieved through the comprehensive analysis of data derived from a multitude of sources, inclusive of logs, metrics, and alerts. The analytical process leverages machine learning and other advanced AI techniques to discern patterns and anomalies that may indicate the root cause of a particular case. The primary objective of RCA within the context of AIOps is to expedite the resolution process of issues so as to reduce Mean Time to Recovery(MTTR) and to preempt their recurrence in the future. This not only enhances operational efficiency but also contributes to the overall reliability and robustness of the systems. Advanced RCA diagnostic methodologies are proposed spanning across academia and industry.

Unlike \cite{sole2017survey}, which explores RCA broadly across various fields, this survey specifically narrows the scope, focusing on approaches that enable a more precise understanding of the nuances and complexities involved in AIOps, especially microservices RCA. It provides a deeper and more targeted analysis of the methodologies, challenges, and future trends related to identifying the root causes of issues within microservices architectures. \cite{xia2022toward, siebert2023applications} primarily focus on graph-based RCA methods. By casting a broader net, this survey examines a wide range of categorical approaches. It captures the diversity of RCA techniques available, offering a more comprehensive understanding of the field and its various methodologies. This survey ensures that readers gain a holistic view of RCA practices in microservices, enabling them to make more informed decisions when selecting or developing RCA strategies. While other reviews might touch on both anomaly detection and RCA \cite{soldani2022anomaly, cheng2023ai}, this survey deliberately focuses solely on RCA. It offers a comprehensive breakdown of different RCA methodologies, delving into their challenges and future trends within microservices environments. This detailed exploration fills a gap in the existing literature, which often overlooks the specificities of RCA in favor of a broader discussion that includes anomaly detection. Incorporating the latest research methodologies, this survey not only reviews historical approaches to RCA but also includes cutting-edge research. At a turning point in the widespread application of advancements in AI and automation, this survey, while reviewing past approaches, provides some possible guidance for subsequent research. 

\begin{figure}[H]
  \centering
  \includegraphics[width=1.5 \textwidth]{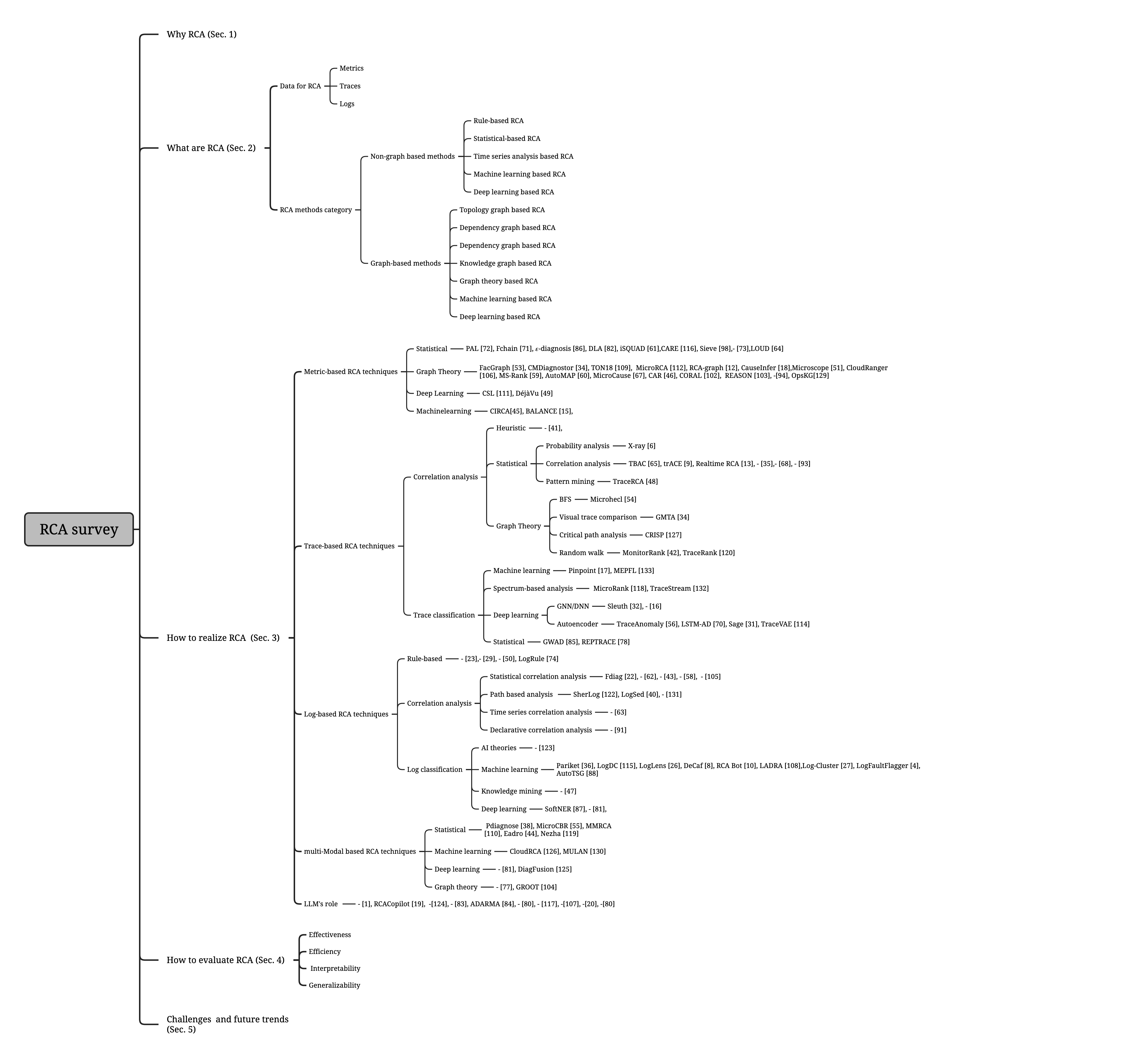}
  \caption{Structure of this paper.}
  \label{fig:RCA}
  \Description{Structure of this paper.}
\end{figure}

The remainder of the paper is organized as follows (shown in Figure ~\ref{fig:RCA} ). Section 2, Preliminaries, introduces the data used for RCA, the category of graphs in RCA, and various RCA methods. Section 3, Methodologies, delves into approaches from five different aspects: metric-based RCA (Section 3.1), trace-based RCA (Section 3.2), log-based RCA (Section 3.3), multi-modal RCA (Section 3.4), and LLM’s role in enhancing RCA (Section 3.5). Each subsection discusses the respective approaches. Section 4 presents an Evaluation of RCA, assessing the performance of these methodologies. Finally, Section 5 explores the broader Challenges and Future Trends in the field of RCA.

\section{Preliminaries}
In the realm of cloud services, an incident signifies any event that disrupts regular service operations or degrades SLA. Upon such incidents arise, a root cause analysis is undertaken to identify the underlying issue responsible for the disturbance. RCA, in the realm of cloud services, represents a sophisticated endeavor involving several stages:
\begin{itemize}
\item {\texttt{Data Collection}}: The initial phase entails gathering relevant data about the incident from various sources, such as logs, metrics (or KPIs), traces, as well as events stemming from publications or configuration changes.
\item {\texttt{Data Analysis}}: Subsequently, the amassed data undergoes thorough analysis to discern patterns, deviations, or connections that may offer insights into the underlying cause of the incident.
\item {\texttt{Hypothesis Validation}}: Following the data analysis, potential root causes are hypothesized and subsequently validated by SREs.
\end{itemize}
\subsection{Data for RCA}
\begin{figure}[h]
  \centering
  \includegraphics[width=0.7 \textwidth]{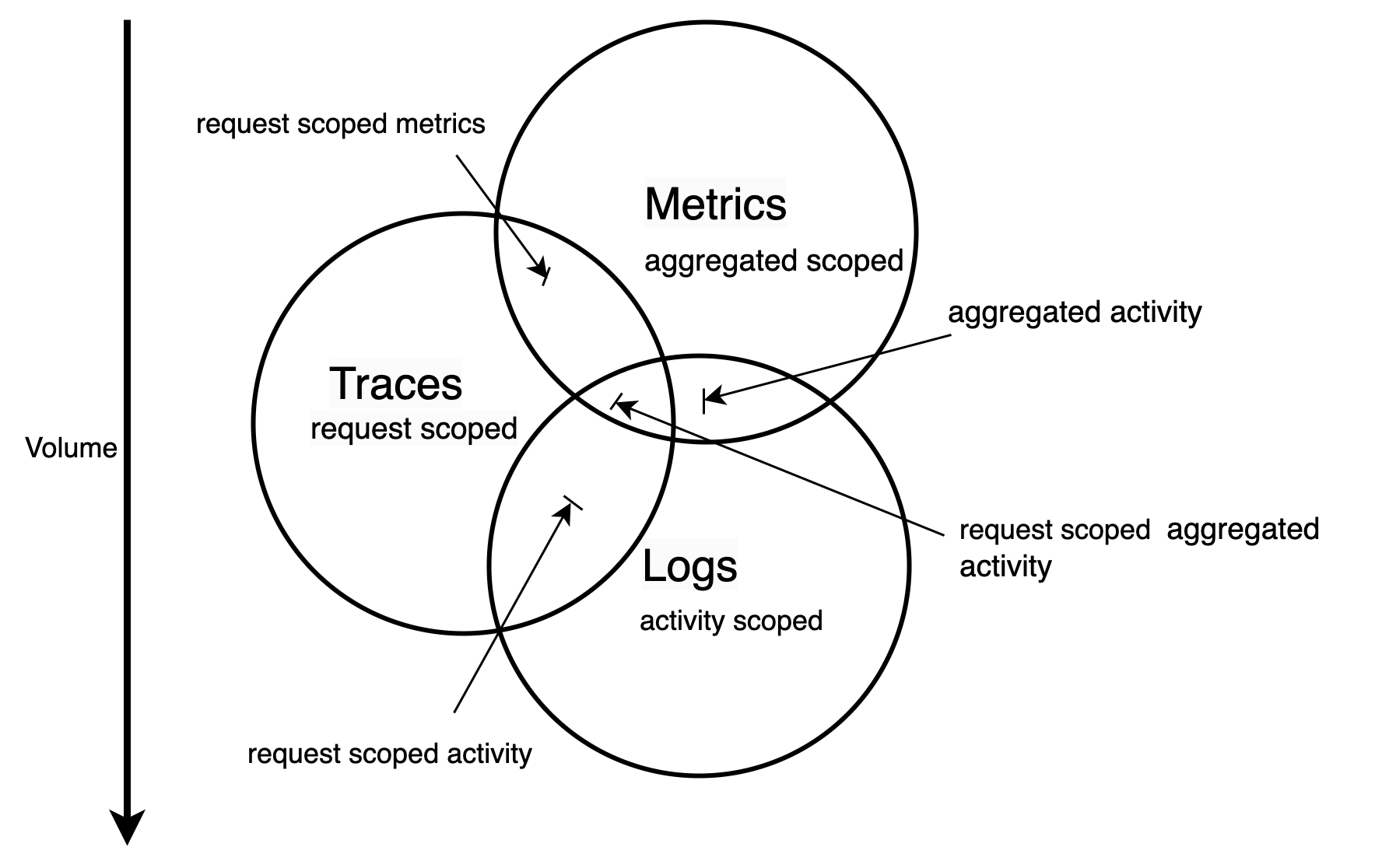}
  \caption{Metrics-traces-logs scope}
  \label{fig:Metrics-traces-logs}
  \Description{Figure 2. Fully described in the text.}
\end{figure}
\begin{figure}[h]
  \centering
  \includegraphics[width=\linewidth]{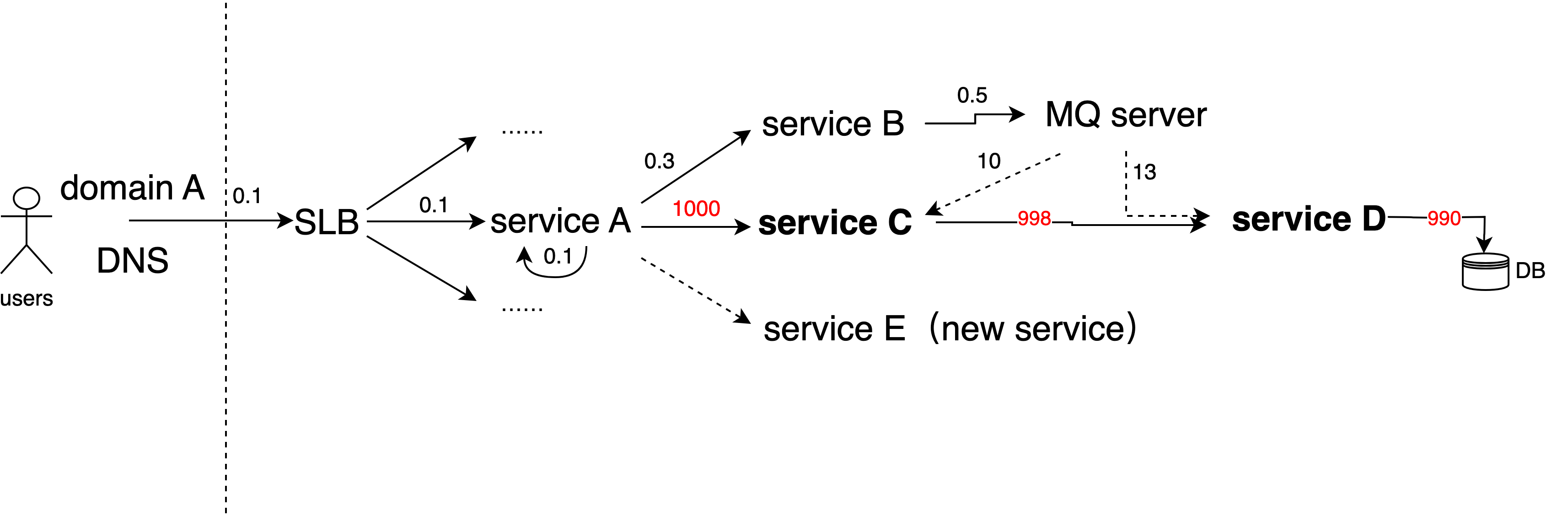}
  \caption{Distributed service request trace}
  \label{fig:traces}
  \Description{Distributed service request trace showing user access to service domain A. The trace illustrates the path taken by the request through multiple services, specifically from service A to C, then to D, and finally to the database. Notably, the total time taken along the path A->C->D->DB was 1000ms, with a significant delay of 990ms observed between D and DB, indicating a possible performance degradation in DB access.}
  \end{figure}

Data for root cause analysis can originate from various sources, including metrics, traces, or logs. When selecting datasets for such analysis, the following types are commonly considered. Figure ~\ref{fig:Metrics-traces-logs} demonstrates the scope, intersection, and volume of metric data, trace data, and log data. Metrics data encompasses system metrics, such as CPU utilization, memory consumption, disk I/O activity, and network traffic, as well as transaction metrics like Queries Per Second (QPS) and Top Percent 99 (TP99). By scrutinizing fluctuations in these performance indicators, deviations from normal system behavior can be identified. Metrics often harbor a substantial amount of anomalous data, providing a direct window into potentially faulty or impacted modules, frequently leveraged for precise fault localization. Nevertheless, the intricacies of call relationships among metric data, graph construction methodologies, and the sequential triggering of metric alerts all play pivotal roles in conducting an accurate root cause analysis. Table \ref{tab:Metric} lists the hierarchical  metric groups of microservice A. Log data, encompassing application, system, and network logs, constitute the predominantly utilized form of data. These logs harbor an abundance of operational status information, encompassing error messages, warning notifications, and status alterations, among others. This diverse array of data serves as a valuable tool for identifying the underlying causes of emerging issues. However, logs often come with significant challenges, given their vast volume, unstructured nature, and diverse formats. Additionally, the rate at which these logs are gathered can be influenced by the sampling rate, which must strike a careful balance between cost-effectiveness and the adequacy of data sampling. Trace data serves as a valuable tool for documenting and analyzing request paths within distributed systems. In intricate microservice architectures, where a single request often traverses multiple services to reach completion, trace data offers critical insights. By meticulously logging the commencement and termination times of each service involved, as well as their interconnected relationships, a comprehensive map of the request's journey can be assembled. Trace data reveals a detailed link topology and call relationships, providing fine-grained information invaluable for graph-based RCA. However, the potentially extensive trace links, resulting from complex call relationships, may undergo truncation, thereby complicating the root cause analysis. Figure ~\ref{fig:traces} demonstrates a distributed service request trace with request timeout anomaly. The trace illustrates the path taken by the request through multiple services, specifically from service A to C, then to D, and finally to the database (DB). Notably, the total time taken along the path A->C->D->DB was 1000ms, with a significant delay of 990ms observed between D and DB, indicating a possible performance degradation in DB access.
\setlength{\tabcolsep}{1pt}
\begin{table}
  \caption{Metric groups of microservice  A }
  \label{tab:Metric}
  \begin{tabular}{ccp{9cm}}
    \toprule
    Category&Group&Metrics\\
    \midrule
App(service) & Transaction & QPS, RT, tp99, log\_error\_count, key business metrics such as total sales, average order value, and conversion rates. \\   
Java & JVM & jvm\_heap\_memory\_usage, jvm\_non\_heap\_memory\_usage, jvm\_thread\_count, jvm\_gc\_frequency, jvm\_gc\_duration, jvm\_class\_loading, young\_gc\_count, full\_gc\_count, young\_generation\_usage, old\_generation\_usage, survivor\_s0\_usage, survivor\_s1\_usage, jvm\_thread\_pool\_waiting\_count \\   
Component & DB & IOPS, connection\_count, queue\_length/waiting\_connections, active\_connections, database\_size, slow\_query\_count, transaction\_throughput, response\_time, error\_rate, session\_durations \\   
Component & Kafka & broker\_connections, topic\_message\_backlog, topic\_consumer\_lag, topic\_consumer\_offset \\   
Container runtime & CPU & container\_cpu\_usage, container\_idle\_cpu, container\_user\_cpu\_time, io\_cpu\_usage \\   
Container runtime & load & 1/5/10\_mins\_container\_average\_load \\   
Container runtime & Memory & container\_memory\_util, container\_memory\_free \\   
Container runtime & Disk & container\_disk\_io\_util, container\_disk\_io\_r, container\_disk\_io\_w \\   
Container runtime & network & container\_network\_in\_rate, container\_network\_out\_rate \\   
K8s & ingress & Used\_space, request\_response\_time, error\_rate, load\_balancer\_util, backend\_health\_checks, TLS\_handshake\_time, caches\_hit\_rate, traffic\_distribution, config\_change\_events \\   
K8s & kubelet & Kubelet\_runtime\_operations\_total, kubelet\_runtime\_erroers\_total \\   
K8s & kube-proxy & connection\_status, active\_connections, idle\_connections, load\_balancing\_status, load\_balancing\_algorithm, abnormal\_exit\_count, restart\_count, failover\_status, request\_duration, throughput, rule\_sync\_status, iptables\_rules\_count/ipvs\_rules\_count \\   
K8s & kuber-scheduler & scheduler\_status, scheduler\_scheduler\_cache\_size, scheduling\_attempts, scheduling\_successes, scheduling\_failures, scheduling\_latency, pending\_pods, scheduled\_pods, available\_nodes \\   
K8s & kube-controller-manager & workqueue\_adds\_total, work\_queue\_depth, controller\_runtime\_reconciler\_queue\_work\_duration, controller\_runtime\_reconciler\_reconcile\_total, controller\_manager\_operation\_duration, controller\_manager\_errors\_total \\   
K8s & kube-apiserver & apiserver\_request\_duration\_seconds, apiserver\_request\_total, apiserver\_request\_errors\_total \\   
Physical machine & CPU & cpu\_usage, idle\_cpu, system\_cpu\_time, user\_cpu\_time, system\_block/wait\_queue\_length, ZombieProcess \\   
Physical machine & Memory & memory\_util, memory\_free, Buffers, cached\_memory, swap\_used, swap\_free, page\_faults, page\_in, page\_out \\   
Physical machine & Disk & disk\_io\_util, disk\_io\_r, disk\_io\_w, disk\_await, disk\_average\_queue\_size, disk\_r\_io\_operation\_per\_sec, disk\_read\_KB\_per\_sec, disk\_average\_service\_time, disk\_w\_io\_per\_sec, disk\_w\_io\_operation\_per\_sec, FS\_max\_avaliable\_space, FS\_max\_util, FS\_total\_space, FS\_used\_percentage, FS\_used\_space \\   
Physical machine & network & network\_in/out\_rate, network\_packet\_loss\_ratio, bandwidth\_consumption, received/sent\_packets, received/sent\_errors\_packets, queue \\   
  \bottomrule
\end{tabular}
\end{table}
  
In addition to harnessing real-world datasets from both public and private sources, the creation of datasets can also be aided by employing diverse benchmarks. Among the numerous benchmarks available for microservices evaluation, several stand out, including TrainTicket \cite{FudanSELab_train_ticket}, Sockshop \cite{helidon_sockshop_sockshop}, OnlineBoutique \cite{lseino_onlineboutique}, and SocialNetwork \cite{opensource_socialnetwork}. These benchmarks play a pivotal role in the evaluation and advancement of microservice architectures by offering standardized environments crucial for conducting RCA.

When conducting root cause analysis, ensuring data reliability is absolutely essential, whether the data comes from metrics, traces or logs. For example, anomalies must be monitorable, which means they should be detectable through metrics. Moreover, these anomalies need to be identifiable through logs, requiring that they are logged with corresponding log levels. Similarly, anomalies should be reflected in traces, indicating that the impacted business processes have undergone comprehensive tracking and instrumentation. In situations where monitoring is not feasible, and anomalies cannot be identified through metrics, traces or logs, even the most advanced root cause analysis methods would be rendered useless. Therefore, ensuring the reliability of data is fundamental to the success of any root cause analysis.
\subsection{RCA Methods}
\begin{figure}[h]
  \centering
  \includegraphics[width=\linewidth]{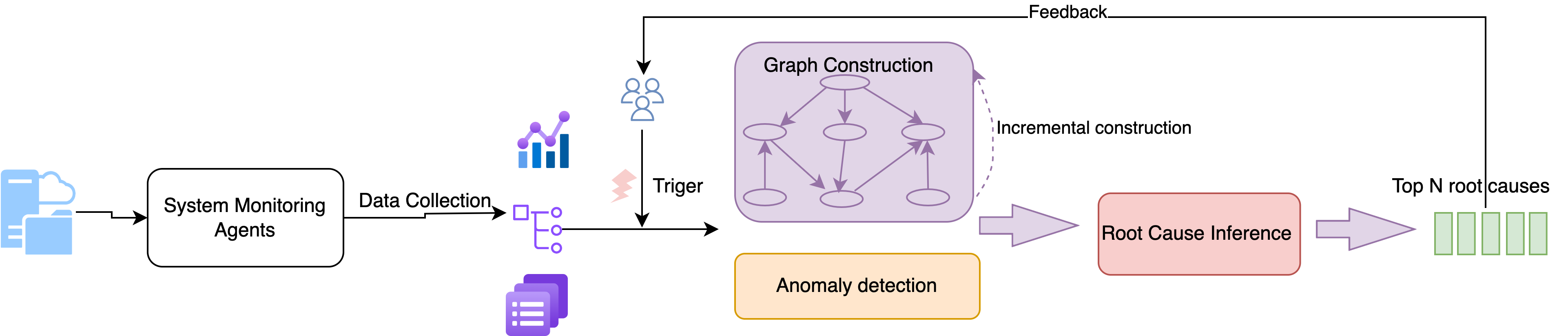}
  \caption{A framework of graph-based RCA}
   \label{fig:graph-rca}
  \Description{A framework of graph-based RCA. This framework outlines a process where system monitoring agents collect various data, including metrics, traces, and log data. This information then undergoes anomaly detection and graph construction. Followed by graph-based RCA, the final output is a structured list of the top N potential root causes for the detected anomalies.}
\end{figure}

The Dependency (Topology) Graph illustrates the intricate interdependencies among various components, modules, services, or systems, which are prevalent entities in root cause analysis. Dependency graph based  approaches involve constructing a graph utilizing Key Performance Indicators (KPIs) and domain expertise. Subsequently, anomalous subgraphs or paths are identified and extracted in response to observed anomalies. The utilization of dependency graphs in root cause analysis aligns seamlessly with actual manual judgment when determining the influence among modules, components, and services. Within dependency graphs, the status of given node can directly impact other nodes that are dependent on it. Dependency graphs serve as a valuable tool for tracing potential problem propagation paths in RCA. For example, in the event of a malfunction in a foundational service, the dependency graph can reveal all upper-tier applications reliant on this service. This, in turn, facilitates the determination of the disruption's scope and potential underlying causes. In contrast to the dependency graph, topology places greater emphasis on the structure of the network, emphasizing the connections between elements rather than their dependency or causal relationships. This approach provides a comprehensive visualization of the network's layout, enabling a more holistic understanding of its components and their interactions. Figure ~\ref{fig:graph-rca} illustrates a framework of graph-based RCA.

Typically, in Bayesian Network graphical model, variables are represented as nodes, and the conditional dependencies among them are depicted by directed edges. The model adheres to the local Markov property, which asserts that any given node is conditionally independent of all its non-descendants when conditioned on its parent nodes. This property ensures a structured and efficient representation of the conditional relationships within the model.  

\begin{figure}[h]
  \centering
  \includegraphics[width=0.7 \textwidth]{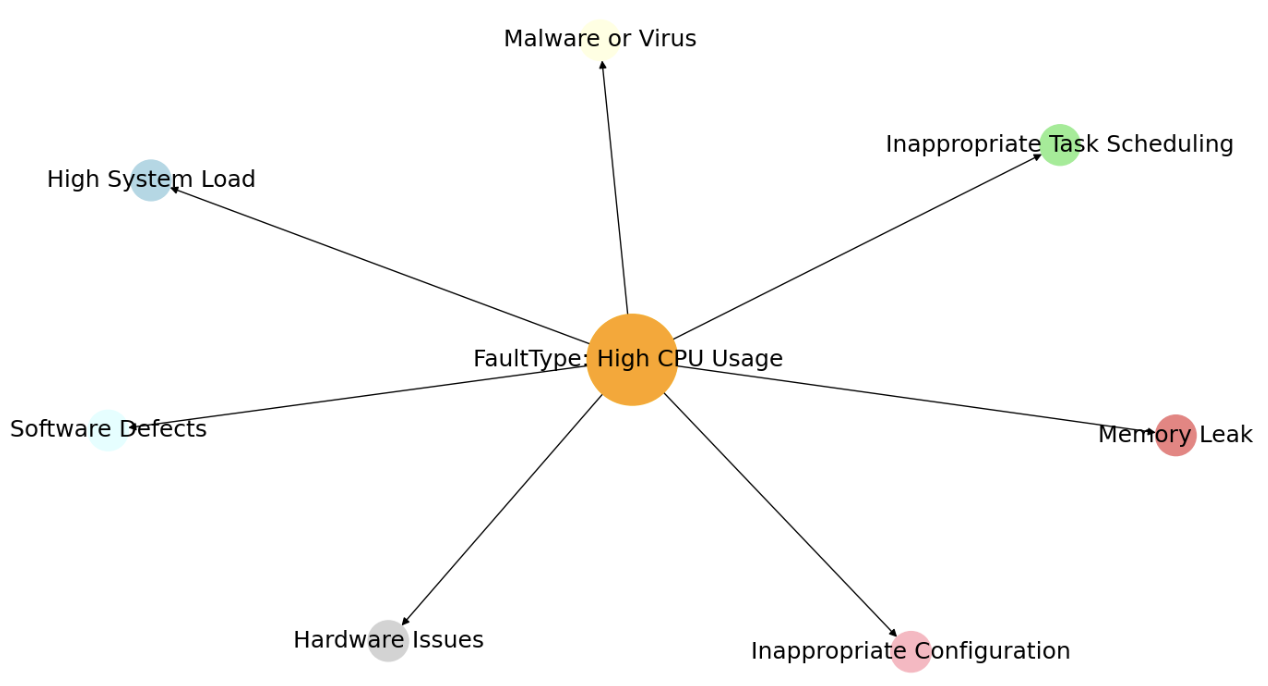}
  \caption{An O\&M knowledge graph of high CPU usage}
   \label{fig:kg-rca}
  \Description{An O\&M knowledge graph of high CPU usage. The graph visually represents the relationships between various factors contributing to high CPU usage, including high CPU load, memory leak, inappropriate task scheduling, and software defects. These nodes are interconnected, indicating their potential causal relationships and impacts on overall system performance.}
  \end{figure}

Operation and Maintenance Knowledge Graph (O\&M Knowledge Graph) offers a way to consolidate and represent the intricate relationships existing between the various components and entities that constitute complex systems. This is accomplished by seamlessly integrating both structured and unstructured data culled from a wide array of sources. Multi-modal integration significantly aids in the comprehension and analysis of the intricate web of interactions and dependencies among system components. Construct a knowledge graph (KG) based on pre-defined entities, relationships, and their associated attributes. Utilize semantic querying and reasoning to retrieve relevant results. This can depict the knowledge graph of either a flawed application architecture or the knowledge graph representing various fault types. The constructed KG, housed in a graph database, allows for swift access and exploration. Furthermore, the semantic nodes and relationships within the KG can be leveraged for causal inference and associative analysis, enhancing the depth and breadth of insights derived from the data. Figure ~\ref{fig:kg-rca} presents  a knowledge graph of high CPU usage case.
  
Rule-Based RCA relies on predefined rules and patterns to identify the root cause of a problem. These methods work well in clear, rule-defined scenarios, but may require significant human involvement and maintenance when dealing with complex, dynamically changing IT environment issues.

Statistical-Based RCA employs statistical techniques to assess the correlation between two Key Performance Indicators (KPIs) by analyzing the relatedness of metrics from historical data, such as the Pearson correlation coefficient or the Spearman rank correlation coefficient, to identify and predict the root cause. This method can handle large amounts of data but may require professional statistical knowledge. The effectiveness may be affected when the data quality is poor or the data volume is insufficient.

Time Series Analysis-Based RCA. In AIOps, operational monitoring metrics and business monitoring metrics exhibit daily, weekly, and other time-series periodicities. Time series analysis can be used to identify abnormal patterns in these data, thereby finding out the root cause of the problem.

Graph Theory-Based RCA. In the microservices environment, complex interrelations exist among services, between services and components, as well as between components and the underlying infrastructure. These interactions are reflected in various graphical representations, including dependency graphs, topological graphs, and causal graphs, which illustrate call dependencies and event causality. By examining the nodes (representing elements) and edges (representing relationships) within these graphs, researchers can conduct root cause analysis. For example, analyzing graph paths and connectivity reveals the patterns of problem propagation, pinpointing the routes through which failures spread. Additionally, centrality measures, such as degree centrality, closeness centrality, and betweenness centrality, enable us to identify the most critical nodes in the network, which are potential root causes of issues. Furthermore, community detection techniques aid in recognizing tightly-knit clusters of nodes, possibly indicating groups of entities sharing similar functionalities or problem areas. By employing graph traversal methods like breadth-first search, we can efficiently locate potential root causes of problems within the microservices architecture.

Machine Learning/Deep Learning-Based RCA. Machine learning and deep learning techniques can be employed to process intricate datasets for the purpose of identifying and predicting root causes, based on the constructed graph structures. As an example, graph neural networks (GNNs) can model the intricate relationships and interdependencies among system components by harnessing the inherent graph-based structure of the data. These networks encapsulate dependencies within graphs through mechanisms like message passing or neighborhood aggregation, thereby facilitating accurate root cause analysis.
\section{Methodologies }
\subsection{Metric-based Root Cause Analysis Techniques}
Metric-based Root Cause Analysis can be categorized into graph-based and non-graph methods. Table \ref{tab:metric} illustrates the investigated root cause analysis approaches of metrics. Among the non-graph methods, PAL \cite{nguyen2011pal} and FChain \cite{nguyen2013fchain} firstly conduct direct KPI correlation analysis. In contrast, PAL derives the propagation pattern within components, whereas FChain takes into consideration both fault propagation patterns and intercomponent dependencies. DLA analyzes the metrics (CPU, Memory, or Network usage) associated with each level (container, node, micro-service) to pinpoint the source of the issue by training a Hierarchical Hidden Markov Model (HHMM) \cite{samir2019dla}. 
\cite{shan2019diagnosis} aims to identify the e correlation in time by measuring the similarity of time series. iSQUAD \cite{ma2020diagnosing} diagnoses cause of intermittent slow Queries (iSQs) by clustering iSQs in databases. \cite{xu2021care} leverages random control trials (RCT) to obtain less ambiguous data in data-driven root cause analysis.

On the other hand, graph-based methods are widely used in RCA. Methods based on Dependency Graphs typically begin by employing a graph construction approach, initially creating causal graphs or topological graphs. Methods \cite{thalheim2017sieve, lin2018facgraph, weng2018root, brandon2020graph, wu2020microrca, guo2020graph, wu2020performance, li2022actionable, li2022causal, chen2023balance} perform root cause analysis on the constructed graph by building topological or dependency graphs. During the dependency graph construction process, Granger Causality \cite{thalheim2017sieve} and PC algorithm \cite{lin2018facgraph, li2022causal} are used to construct completed partially directed acyclic graph. There are also more through the trace request to construct the dependency graph \cite{weng2018root, brandon2020graph, wu2020microrca, wu2020performance, chen2023balance}, there are also through the law of large numbers and the Markov properties \cite{guo2020graph}, deployment relationships to construct call graph or dependency graph \cite{li2022actionable}. Quite many approaches use metric data to construct call graphs, however, ambiguous correspondences between upstream and downstream calls may exist and result in exploring unexpected edges in the constructed call graph, while advocate conducting RCA on this graph may lead to misjudgments \cite{yu2023cmdiagnostor}. CMDiagnostor leverages the law of large numbers and the Markov properties of network traffic to investigate ambiguity problem and construct an ambiguity-free call graph.

Once the dependency graph has been constructed, subsequent RCA can be performed through various methodologies, including graph theory, machine learning, and deep learning. These approaches leverage the established graph to analyze data for fault diagnosis and problem resolution. \cite{thalheim2017sieve} uses correlation of a correct version and faulty version to realize RCA. FacGraph \cite{lin2018facgraph} and RCA-graph \cite{guo2020graph} utilize subgraph mining (FSM) algorithm graph or graph similarity-based pattern matching through traversal techniques, such as breadth-first search (BFS) ordered strings and depth-first search (DFS) on dependency graphs, to identify the root cause. TON18 \cite{weng2018root}, MicroRCA \cite{wu2020microrca} and \cite{sun2021fault} employ random walks on graphs for root cause analysis. Concurrently, \cite{brandon2020graph} adopts subgraph similarity pattern matching on topological structures for fault diagnosis. 

The advancement of deep learning methods on dependency graphs has led to significant progress in fault diagnosis. \cite{wu2020performance} applies an autoencoder to identify abnormal service metrics, utilizing a ranked list of reconstruction errors to pinpoint potential culprit services. On the other hand, D\'ej\`aVu \cite{li2022actionable} methodology harnesses Gated Recurrent Units (GRU) and Convolutional Neural Networks (CNN) to encode the metrics of a fault unit into a fixed-length vector. It then employs a binary classification scheme to output a root cause score ranging from 0 to 1 for each fault unit, representing the likelihood of the unit being the root cause of the failure, based on the aggregated features. Furthermore, D\'ej\`aVu posits that recurrent events are of particular interest RCA, as they provide valuable insights into recurring patterns of failure.
\setlength{\tabcolsep}{0pt}
\begin{table*}  
  \caption{Investigated root cause analysis approaches of metrics}  
  \label{tab:metric}  
  \begin{tabular}{cccp{3.73cm}cc} 
    \toprule  
    Method & Data source & RCA category & RCA method & Graph construction & Graph category \\  
    \midrule  
    PAL \cite{nguyen2011pal} & Metrics & Statistical & Direct KPI correlation & - & Non-graph \\  
     Fchain \cite{nguyen2013fchain} & Metrics & Statistical & Direct KPI correlation & - & Non-graph \\  
\(\varepsilon\)-diagnosis \cite{shan2019diagnosis} & Metrics & Statistical & Similarity of time series & - & Non-graph \\  
DLA \cite{samir2019dla}& Metrics & Statistical & Hierarchical Hidden Markov Model analysis & - & Non-graph \\  
iSQUAD \cite{ma2020diagnosing} & Metrics & Statistical & Type-Oriented Pattern Integration Clustering & - & Non-graph \\  
CARE \cite{xu2021care} & KPIs & Statistical & Random Control Trial & - & Non-graph \\  
Sieve \cite{thalheim2017sieve} & Metrics & Statistical & KPI correlation& Granger causality & Dependency graph \\  
FacGraph \cite{lin2018facgraph} & Metrics & Graph Theory & Breadth first ordered string & PC algorithm & Topology graph \\  
CMDiagnostor \cite{guo2020graph} & Metrics & Graph Theory & DFS on dependency graph & BFS & Dependency graph \\  
TON18 \cite{weng2018root} & Metrics & Graph Theory & Random walk on dependency graph & Trace request & Dependency graph \\  
MicroRCA \cite{wu2020microrca} & Metrics & Graph Theory & Random walk on topology graphs & Trace request & Dependency graph\\  
RCA-graph \cite{brandon2020graph} & Metrics & Graph Theory & Subgraph matching on topology & Trace request & Topology graph \\  
CSL \cite{wu2020performance} & Metrics & Deep Learning & Autoencoder & Trace request & Dependency graph \\  
D\'ej\`aVu \cite{li2022actionable} & Metris, events & Deep Learning & Deep learning on dependency graph & Deployment structure & Dependency graph \\  
CIRCA \cite{li2022causal} & Metrics & Machine learning & Bayesian network & System knowledge & Dependency graph \\  
BALANCE \cite{chen2023balance} & KPIs & Machine learning & Bayesian network & Trace request & Dependency graph \\  
- \cite{nie2016mining} & Metrics,events & Statistical & Correlation of causality graph & Monitoring data & Causality graph \\  
LOUD \cite{mariani2018localizing} & KPIs & Statistical & Correlation of causality graph & Trace request & Causality graph \\  
CauseInfer \cite{chen2014causeinfer} & Metrics & Graph Theory & DFS on causality graph & PC algorithm & Causality graph \\  
Microscope \cite{lin2018microscope} & Metrics & Graph Theory & Reverse DFS on causality graphs & PC algorithm & Causality graph \\  
CloudRanger \cite{wang2018cloudranger} & Metrics & Graph Theory & Random walk on causality graph & PC algorithm & Causality graph \\  
MS-Rank \cite{ma2019ms} & Metrics & Graph Theory & Random walk on causality graph & PC algorithm & Causality graph \\  
AutoMAP \cite{ma2020automap} & Metrics & Graph Theory & Random walk on causality graph & PC algorithm & Causality graph \\  
MicroCause \cite{meng2020localizing} & Metrics & Graph Theory & Random walk on causality graph and path  \newline condition time series & PC algorithm & Causality graph \\  
CAR \cite{li2022mining} & Metrics & Graph Theory  & random walk on causality graph & FPG-Miner & Causality graph \\  
CORAL \cite{wang2023incremental} & Metrics & Graph Theory & Random walk on causality graph & Customized method & Causality graph \\  
REASON \cite{wang2023hierarchical} & Metrics & Graph Theory & Random walk on causality graph & GNN & Causality graph \\  
CausIL \cite{chakraborty2023causil} & Metrics & - & - & Customized method & Causality graph \\  
- \cite{sun2021fault} & Multiple factors & Graph Theory & Random walk on dependency graph & Customized method & Knowledge graph \\  
OpsKG \cite{zhao2023design} & Metrics & Graph Theory & MicroRCA & NLP & Knowledge graph \\  
    \bottomrule  
  \end{tabular}  
\end{table*}

As a classic example of dependency graphs, Bayesian Networks have garnered attention from both researchers and industry practitioners for it excel in probabilistic deduction and causal inference, particularly when dealing with uncertain and incomplete information. CIRCA \cite{li2022causal} takes full advantage of the Causal Bayesian Network (CBN), bridging between observational data and interventional knowledge, the change of probability distribution conditioned on the parents in the CBN. \cite{chen2023balance} exploits the explainable AI (XAI) Bayesian Linear Attribution (BALANCE) for RCA. BALANCE uses a Bayesian multi collinear feature selection (BMFS) model to predict the target KPIs given the candidate root causes in a forward manner while promoting sparsity and concurrently paying attention to the correlation between the candidate root causes. And it introduces attribution analysis to compute the attribution score for each candidate in a backward manner. 

Unlike dependency or topology aimed at understanding the structure and connectivity of a system, while causal graph is concerned with the cause-and-effect relationships within the system. In particular, causal graph are instrumental in identifying and interpreting the underlying reasons for RCA. Causality Graph-based RCA also comprises two stages: the initial stage involves generating a directed acyclic the causality graph with spatiotemporal restriction, followed by performing RCA on the causality graph. In the realm of causal relationship discovery, considerable number of researchers \cite{chen2014causeinfer, lin2018microscope, wang2018cloudranger, ma2019ms, ma2020automap, meng2020localizing} have also constructed component or service causality graphs using the PC algorithm and its variants and they put more emphasis on RCA methods. \cite{nie2016mining} employs a data mining method to extract an initial causality graph, which is then refined using the random forest algorithm to construct a more precise causality graph. The correctness and completeness of causality discovery influents second stage RCA. Concerning this, CAR \cite{li2022mining} discovers almost a complete graph based on XG-Boost active learning in case of miss a large proportion of relations and fluctuation propagation among time series. REASON \cite{wang2023hierarchical} employs novel hierarchical graph neural networks to model both intra-level and inter-level non-linear causal relations, thereby constructing interdependent causal networks. CORAL \cite{wang2023incremental}, an incremental disentangled causal graph learning approach, aims at decoupling state-invariant and state-dependent information. Completeness and complexity often present a paradox in causality graph analysis, where intricate causal relationships can lead to reduced efficiency and accuracy in processing. CausIL \cite{chakraborty2023causil} provides a list of general rules in the form of prohibited edges, along with ways to incorporate them, aiming to improve computation time and accuracy. 

Following the construction of the causality graph, in the second phase of causality graph analysis, correlation of causality graph analysis is employed for conducting root cause analysis \cite{nie2016mining, mariani2018localizing} under the hypothesis that the anomalous KPIs related to a fault are highly correlated and form a connected subgraph of the propagation graph. \cite{chen2014causeinfer, lin2018microscope}traverse the causal graph in DFS (Depth First Search) and reverse DFS to inspect the neighboring nodes of each node, introducing correlation coefficient to measure the relationship between the service level objective of the anomalous front service and that of the potential root cause candidates. Numerous researchers tend to utilizes random walks on the constructed causality graph, and place greater emphasis on the construction of the causality graph itself. CloudRanger \cite{wang2018cloudranger}, MS-Rank \cite{ma2019ms}, AutoMAP \cite{ma2020automap} MicroCause \cite{meng2020localizing}, CAR \cite{li2022mining}, CORAL \cite{wang2023incremental}, and REASON \cite{wang2023hierarchical} apply random walk to perform RCA leveraging the learned interdependent causal networks. MicroCause captures the sequential relationships in time series data when performing temporal cause oriented random walk.

Knowledge graph within the domain of expertise is referred to as O\&M Knowledge Graph, a completely undirected graph. In O\&M Knowledge Graph, the entity and relation demonstrate a faulty architecture. \cite{sun2021fault, saha2022mining} and OpsKG \cite{zhao2023design} construct O\&M knowledge graph by extracting a joint entity-relation extraction according to alarm query, root-cause analysis and location, and alarm prediction. And a random walk fault path lookup approach in the root cause localization stage. OpsKG adopts the MicroRCA \cite{wu2020microrca} method into OpsKG to implement RCA, which, in essence, still utilizes random walk.
\subsection{Trace-based Root Cause Analysis Techniques}
Trace-based RCA employs traces to pinpoint the fundamental reasons behind issues by scrutinizing the sequence of events that preceded the occurrence of the problem. Different with metric-based RCA, trace-based RCA centers on traces instead of metrics or logs. Since traces naturally exhibit invocation relationships, trace-based techniques are inherently associated with dependency or topology graphs. However, these dependency graphs, which may encompass temporal or regular trace dependencies, may not possess the completeness of a comprehensive dependency graph. Among dependency graph-based methods, statistical analysis and classification on traces are primarily employed. Table \ref{tab:trace} illustrates investigated root cause analysis approaches of traces.
\subsubsection{Correlation Analysis}
Non-graphical methods are not frequently used in trace RCA. \cite{julisch2003clustering} leverages alarm clustering techniques to support the analysis process by identifying and eliminating the most predominant and persistent root causes of alarms. X-ray \cite{attariyan2012x} employs dynamic flow analysis to associate events with potential root causes, prioritized by probability. The aggregate cost for each root cause is calculated by summing the product of individual event costs and their likelihood of being caused by a specific root cause. 

Dependency graph constructed from monitoring data may exhibit incompleteness, as not all potential interactions among architectural elements are necessarily represented by edges in the graph \cite{marwede2009automatic, 2003Discovering}. Drawing from reconstruction of architecture models from monitoring data and the subsequent evaluation of anomaly detectors, the caller-callee relationships, along with their respective anomaly scores, can be discerned. \cite{marwede2009automatic} introduces the anomaly correlator, RanCorr, a set of rules designed to scrutinize calling dependencies among software components within the constructed graph. The correlation analysis is facilitated by implementing a rule set founded on general assumptions about propagation. Ultimately, the visualization of potentially correlated anomalies serves to bolster decision-making. \cite{cai2019real} utilize a heuristic approach to compare suspicious traces with predefined normative patterns, which relies on graph properties of sequential patterns, flagging anomalies when deviations exceed the 95\% statistical confidence limits. They employ a regression-based relative importance analysis, which assesses the explained variance of the total response time in relation to the latency of individual transitions within a trace. TraceRCA \cite{li2021practical} employs frequent pattern mining techniques to narrow down the search space, and ranks microservices with less normal traces passing through. \cite{liu2021microhecl} performs anomaly entity node analysis and propagation chain analysis by traversing the graph along anomalous service call edges. Candidate root causes are then ranked based on the absolute value of the Pearson correlation coefficient. trACE \cite{behera2023trace} pinpoints the root cause within a microservices architecture by exploiting its hierarchical deployment structure. It analyzes dependencies correlation between anomalous Pods and services to infer the causes of service anomalies. While this method excels at identifying root causes at both the node and service levels, it does not directly reveal the triggers behind these failures.

REPTRACE \cite{ren2019root} employs categorizing trace similarity for RCA. They construct a dependency graph based on read/write and parent/child process dependencies, tracing back from inconsistent artifacts to identify the responsible processes. To address the challenges posed by noisy dependencies and uncertain parent/child relationships, REPTRACE focuses on system calls that produce differing outputs across builds and computes the similarity of runtime values to establish relevant dependencies. GMTA \cite{guo2020graph} employs a graph representation to visualize trace data in microservice systems, which allow operators to manually determine the possible root causes. GMTA supports a wide range of analytical applications, including visualizing service dependencies, guiding architectural decisions, analyzing service behavior changes, detecting performance issues, and enhancing the efficiency of problem identification.

Dependencies path analysis microservices is another correlation analysis method \cite{soualhia2022automated, liu2020unsupervised, zhang2022crisp}. \cite{soualhia2022automated} computes potential dependencies among anomaly elements using a correlation model, identifying the critical path leading to the detected fault through this correlation model. TraceAnomaly \cite{liu2020unsupervised} employs a variational autoencoder (VAE) to identify anomalies within traces, utilizes the three-sigma rule to determine anomaly intervals, and defines the root cause as the longest continuous path encompassing these anomalous intervals. Similar to TraceAnomaly, CRISP \cite{zhang2022crisp} uses an anomaly detection framework but encodes only the critical paths into vectors generated through Critical Path Analysis (CPA). CRISP's tracing facility captures the sequence of operations and interactions among system components, using this data to construct a dependency graph. This graph illustrates the relationships between microservices or system elements, allowing CRISP to pinpoint the most influential critical paths for system performance. Through the analysis of these paths, CRISP detects anomalies and bottlenecks, aiding in the diagnosis and resolution of issues in intricate software systems. 

Researchers utilize random walks for path analysis in dependency graph identifying the underlying causes of issues. MonitorRank \cite{kim2013root} advocates that a basic call graph might not accurately represent the true dependency relationships. By harnessing both historical and real-time data from various sensors, the system utilizes a specially designed random walk algorithm alongside a sensor dependency graph. TraceRank \cite{yu2023tracerank} also gathers service invocation data in a microservice setting to assemble a service dependency graph. A PageRank-inspired random walk algorithm is then employed to accurately identify the primary culprits behind any issues. 
\subsubsection{Trace Classification}
Another category for trace RCA is trace classification. Pinpoint \cite{chen2002pinpoint} adopts data mining techniques to cluster client requests based on their successes and failures, correlating them with the components that served them. This correlation aids in pinpointing which components are most probably malfunctioning. MEPFL, similarly harnesses machine learning to predict potential errors, fault locations, and types. This method selects relevant features, preprocesses data, and trains models using Random Forests, K-Nearest Neighbors, and Multi-Layer Perceptron for within a microservices architecture \cite{zhou2019latent}.  \cite{chen2021trace} proposes the use of a supervised learning model, specifically Deep Neural Networks (DNNs), to implement a classifier. This classifier is trained on both normal and faulty datasets to learn patterns associated with both normal and abnormal behaviors, thereby aiding in fault localization. On the other hand, \cite{mi2012localizing} categorizes requests based on call sequences, identify abnormal requests through principal component analysis, and then single out anomalous methods using the Mann-Whitney hypothesis test. They asses the behavioral similarities of all replicated instances of the anomalous methods with the Jensen-Shannon divergence and select those with differing behaviors as the ultimate culprits of performance anomalies.
\setlength{\tabcolsep}{0.5pt}
\begin{table*}  
  \caption{Investigated root cause analysis approaches of traces}  
  \label{tab:trace}  
  \begin{tabular}{cclp{2.7cm}cc} 
    \toprule  
    Method & Data source & RCA category & RCA method & Graph construction & Graph category \\  
    \midrule  
    - \cite{julisch2003clustering} & Traces & Heuristic & Alarms clustering & - & Non-graph \\  
X-ray \cite{attariyan2012x} & Traces & Statistical & Probability analysis & - & Non-graph \\  
- \cite{2003Discovering} & Traces & Statistical & Correlation analysis & Invocation relationship & Dependency graph \\  
TBAC \cite{marwede2009automatic} & Traces & Statistical & Correlation analysis & Invocation relationship & Dependency graph \\  
Realtime RCA \cite{cai2019real} & Traces & Statistical & Regression-based relative analysis & Invocation relationship & Dependency graph \\  
TraceRCA \cite{li2021practical} & Traces & Statistical & Pattern mining with less normal traces passing through & Invocation relationship & Dependency graph \\  
- \cite{mi2012localizing} & Traces,logs & Statistical & Jensen-Shannon divergence analysis & Invocation relationship & Dependency graph \\  
Microhecl \cite{liu2021microhecl} & Traces & Graph Theory & BFS on dependency graphs & Deployment structure & Dependency graph \\  
trACE \cite{behera2023trace} & Traces & Statistical & Correlation Analysis & Deployment structure & Dependency graph \\  
GWAD \cite{setiawan2020gwad} & Traces,events & Statistical & Jaccard's similarity trace classification & Greedy workflow graph & Dependency graph \\  
REPTRACE \cite{ren2019root} & Traces & Statistical & Sequential patterns similarity comparision & Invocation relationship & Dependency graph \\  
GMTA \cite{guo2020graph} & Traces & Graph Theory & Visual trace comparison & Invocation relationship & Dependency graph \\  
- \cite{soualhia2022automated} & Traces & Statistical & Path correlation analysis & Invocation relationship & Dependency graph \\  
CRISP \cite{zhang2022crisp} & Traces & Graph Theory & Critical path analysis & Invocation relationship & Dependency graph \\  
MonitorRank \cite{kim2013root} & Metrics, traces & Graph Theory & Random walk on dependance graphs & Invocation relationship & Dependency graph \\  
TraceRank \cite{yu2023tracerank} & Traces & Graph Theory & Random walk on dependency graph & Invocation relationship & Dependency graph \\  
Pinpoint \cite{chen2002pinpoint} & Traces & Machine learning & Clustering & Invocation relationship & Dependency graph \\  
MEPFL \cite{zhou2019latent} & Traces & Machine learning & Random forest, k-NN, Multi-Layer Perceptron for classification & Invocation relationship & Dependency graph \\  
- \cite{chen2021trace} & Traces & Deep learning & DNN for classification & Invocation relationship & Dependency graph \\  
MircoRank \cite{2021MicroRank} & Traces & Machine learning & Spectrum analysis & Invocation relationship & Dependency graph \\  
TraceStream \cite{zhou2023tracestream} & Traces & Machine learning & Spectrum analysis & Invocation relationship & Dependency graph \\  
Sleuth \cite{gan2023sleuth} & Traces & Deep learning & GNN & Invocation relationship & Dependency graph \\  
TraceAnomaly \cite{liu2020unsupervised} & Traces & Deep learning & Autoencoder & Invocation relationship & Dependency graph \\  
LSTM-AD \cite{nedelkoski2019anomaly} & Traces & Deep learning & Long short-term memory (LSTM) VAE & Invocation relationship & Dependency graph \\  
Sage \cite{gan2021sage} & Traces & Deep learning & Graphical variational autoencoder correlation analysis & Causal Bayesian Networks & Dependency graph \\  
TraceVAE \cite{xie2023unsupervised} & Traces & Deep learning & dual-variable graph VAE & Invocation relationship & Dependency graph \\
    \bottomrule  
  \end{tabular}  
\end{table*}

Spectrum-based analysis is also utilized for trace classification. MicroRank \cite{2021MicroRank} utilizes clues extracted from both normal and abnormal traces to pinpoint the root causes of latency issues. Its PageRank Scorer module employs abnormal and normal trace data, distinguishing the significance of various traces through advanced spectrum techniques. Spectrum method computes a ranking list, leveraging the weighted spectrum data from the PageRank Scorer to identify the root causes. Utilizing weighted spectrum analysis, TraceStream \cite{zhou2023tracestream} generates a vector representation for each trace by extracting key structural and temporal features. Leveraging these representations, traces exhibiting similar behaviors are continuously grouped into normal and anomalous clusters by calculating the centrality of each candidate node and ranking service nodes based on centrality-derived spectrum scores. Sleuth \cite{gan2023sleuth} leverages a graph neural network (GNN) to capture the causal impact of each span in a trace, and trace clustering using a trace distance metric to reduce the amount of traces required for root cause localization.

Autoencoders are also extensively employed in trace root cause analysis, where they are used to learn efficient coding of the input data, enabling the identification of underlying patterns and anomalies that contribute to the occurrence of errors or faults within a system. LSTM-AD \cite{nedelkoski2019anomaly} extracts operation sequences and latency time series from traces, modeling the temporal features using a multi-modal Long Short-term Memory (LSTM) VAE. Uniquely, this method maximizes node anomaly scores instead of summing them up. TraceAnomaly \cite{liu2020unsupervised} employs a variational autoencoder (VAE) to identify anomalies within traces. It utilizes the three-sigma rule to determine anomaly intervals, and defines the root cause as the longest continuous path encompassing these anomalous intervals. Sage \cite{gan2021sage} constructs a visual representation of these complex dependencies utilizing Causal Bayesian Networks, presenting a clear and structured method to decipher the intricate relationships within the microservice architecture. Sage employs counterfactuals via a graphical Variational Autoencoder to simulate hypothetical scenarios and evaluate their potential impact on the cloud service's QoS. Sage not only identifies the underlying issues but also takes proactive steps to restore and maintain optimal QoS levels in cloud services. TraceVAE \cite{xie2023unsupervised} introduces an innovative dual-variable graph variational autoencoder, targeting two types of anomalies commonly found in traces: structural and time-consumption anomalies. TraceVAE encodes the trace structure and utilizes the negative log-likelihood (NLL) as a quantifiable measure of the anomaly score. NLL signifies how improbable it is for that trace to align with the model's learned distribution. 
\subsection{Log-based Root Cause Analysis Techniques}
Log based RCA begins with the acquisition of log data from the system, followed by preprocessing steps such as log parsing, formatting, and noise reduction. Log analysis techniques are applied to detect anomalies within the logs, common techniques including anomaly detection, correlation analysis, pattern mining, and machine learning methods. These techniques specifically involve identifying abnormal patterns in logs, analyzing correlations and causal relationships between log events, mining frequent patterns or behaviors in logs, and classifying and clustering logs. With the assistance of system context information (such as system topology, configuration, metadata, etc.) and knowledge graphs, contextual relationships are integrated. Subsequently, RCA methods are employed to locate the root cause of issues within the logs. Here, RCA methods are categorized into the following types: rule-based RCA, statistical RCA, graph-based RCA, and machine learning-assisted RCA. Table \ref{tab:log} demonstrates the investigated root cause analysis approaches of logs. 

\subsubsection{Correlation Analysis}
Rule-based RCA relies on predefined rules and patterns to identify the root cause of a problem. These methods work well in clear, rule-defined scenarios, but may require significant human involvement and maintenance when dealing with complex, dynamically changing IT environment issues. \cite{cinque2012event, fu2014digging} propose a rule-based approach for failures through event logs analysis. \cite{fu2014digging} clusters frequent event sequences into event groups and extract failure rules based on events of the same types. \cite{lin2020fast, notaro2023logrule} both adopt association rule mining (ARM) to automate RCA. Lin2020fast incorporate a new rule refinement algorithm to reduce the number of redundant explanations in the result set. They use the Apriori and FP-Growth, association rule learning, to find item-sets in structured logs that are strongly linked to association failures.  Built upon this work, LogRule \cite{notaro2023logrule} introduces new compression and rule selection techniques, by removing the need for manual tuning of metric thresholds, and by introducing the use of operators designed specifically for each data types, efficiently preserving semantic information.

Correlation analysis-based methods leverage statistical correlation to identify and analyze relationships between log entries, events in system logs and faults metrics to pinpoint the root causes of failures. Pinpointing the underlying causes of a given failure from extensive logs, \cite{chuah2010diagnosing} reconstructs event sequences and establish connections among events and \cite{tak2016logan} identifies log entries, both of them employs statistical directly correlation analysis. \cite{lu2017log} involves selecting features from structured logs related to execution time, data locality of each task, memory usage, and garbage collection of each node. By analyzing weighted features, they determine the probability of various root causes.To identify root causes, \cite{maeyens2020process} analyzes process on event logs, variations in trace durations, event timings, and path frequency. \cite{koyama2023log} leverages Istio’s default log count computes an indicator, representing the structure of the HTTP response body and logged as an additional field, to identify application errors. It tallies items like maps or lists in JSON format extracted from the HTTP response body, serving as the indicator for error identification. In methods of correlation analysis, \cite{wang2020root} applies correlation metrics to analyze log data following log anomaly detection with DeepLog algorithm.

In the realm of correlation analysis methods, path-based analyses also exist. These approaches are particularly useful when examining the causal pathways that link various factors and outcomes in a complex system \cite{yuan2010sherlog, jia2017logsed, zhou2015distance}. SherLog reports pathways connecting the longest subsequence of logging messages involving the error message. In cases where multiple paths link the same sequence of Logging Points, SherLog compares these paths and initially reports the common records along these paths as the Must-path before presenting the rest \cite{yuan2010sherlog}. LogSed mines a time-weighted control flow graph (TCFG) to represent healthy execution flows in cloud components. By analyzing deviations from this TCFG, LogSed automatically raises anomaly alerts. \cite{zhou2015distance} introduces a distance-based approach calculating distance-based score from call trees.

In the chronological sequence of events, some scholars posit that the antecedent event is the prospective root cause. \cite{mandal2021localization} perform the short-term time series analysis and convert logs into multivariate time series data by counting error logs within specific time bins for affected services. They employ the personalized PageRank to provide a ranked order of potentially faulty nodes. In contrast to approaches that correlate the performance of various service instances or logged events to prioritize root causes based on their likelihood of causing observed failures, without elucidating how these failures propagate to other service instances, \cite{soldani2022failure} offers a fresh perspective. They present a declarative root cause analysis technique capable of determining the potential cascading failures that may have precipitated an observed failure, specifically, identifying the service instances that could have failed initially. 
\setlength{\tabcolsep}{0.4pt}
\begin{table*}  
  \caption{Investigated root cause analysis approaches of logs}  
  \label{tab:log}  
  \begin{tabular}{cccp{3.9cm}cc} 
    \toprule  
    Method & Data source & RCA category & RCA method & Graph construction & Graph category \\  
    \midrule  
- \cite{cinque2012event} & Logs,events & Rule-based  & - & - & Non-graph \\  
- \cite{fu2014digging} & Logs,events & Rule-based  & - & - & Non-graph \\  
- \cite{lin2020fast} & Logs & Rule-based  & Association rule mining & - & Non-graph \\  
LogRule \cite{notaro2023logrule} & Logs, events & Rule-based & Association rule mining & - & Non-graph \\
Fdiag \cite{chuah2010diagnosing} & Logs,events & Statistical & Log entries statistical correlation analysis & - & Non-graph \\  
LOGAN \cite{tak2016logan} & Logs & Statistical  & Log entries correlation analysis  & - & Non-graph \\  
- \cite{lu2017log} & Logs & Statistical  & Weighted features analysis & - & Non-graph \\  
- \cite{maeyens2020process} & Logs,events & Statistical & Event log correlation analysis  & - & Non-graph \\  
- \cite{koyama2023log} & Logs & Statistical & Indicator correlation analysis  & - & Non-graph \\  
- \cite{wang2020root} & Logs & Statistical & Applying correlation metrics to analyze log data & - & Non-graph \\  
SherLog \cite{yuan2010sherlog} & Logs & Statistical & Path correlation analysis  & - & Non-graph \\  
LogSed \cite{jia2017logsed} & Logs & Statistical & Path analysis & Transaction log & Dependency graph \\  
- \cite{zhou2015distance} & Logs & Statistical & Distance-based correlation analysis  & Transaction log & Call graph \\  
- \cite{mandal2021localization} & Logs & Statistical & Time series correlation  analysis  & Granger causality & Dependency graph \\  
- \cite{soldani2022failure} & Logs & Statistical & Declarative  correlation analysis  & - & Non-graph \\  
- \cite{Zawawy2010log} & Logs & AI theories & SAT Solver & - & Non-graph \\  
Pariket \cite{gupta2015pariket} & Logs,events & Machine learning & Decision tree classification & - & Non-graph \\  
LogDC \cite{xu2017logdc} & Logs & Machine learning & Log features clustering & - & Non-graph \\  
LogLens \cite{debnath2018loglens} & Logs & Machine learning & Similarity distance Clustering & - & Non-graph \\  
DeCaf \cite{2019DeCaf} & Logs & Machine learning & Random forest correlation analysis & - & Non-graph \\  
RCA Bot \cite{behera2023root} & Logs & Machine learning & Logistic regression, naïve Bayes, decision trees, and random forests classfication & - & Non-graph \\  
- \cite{suriadi2013root} & Logs,events & Statistical & Log attributes classfication & - & Non-graph \\  
HWLogAnalysis \cite{tan2017hwlog} & Logs,events & Statistical& Two-dimensional frequent pattern mining & - & Non-graph \\  
LADRA \cite{wanga2018ladra} & Logs & Machine learning & General regression neural network likelihood analysis & - & Non-graph \\  
- \cite{ikeuchi2018root} & Logs & Statistical & Repair knowledge model correlation analysis  & - & Non-graph \\  
SBLD \cite{rosenberg2020spectrum} & Logs,events & Machine learning & Spectrum analysis & - & Non-graph \\  
Log-Cluster \cite{ding2014mining} & Logs & Machine learning & Log sequences clustering & - & Non-graph \\  
LogFaultFlagger \cite{lin2016log} & Logs & Machine learning & TF-IDF+KNN log classfication & - & Non-graph \\  
- \cite{amar2019mining} & Logs & Machine learning & Markov decision model & - & Non-graph \\  
- \cite{li2022abc} & Logs & NLP  & Similarities  analysis  & - & Non-graph \\  
SoftNER \cite{shetty2022softner} & Logs & Deep learning & BiLSTM+CRF+Attention Mechanism & NLP  & Knowledge graph \\  
- \cite{saha2022mining} & traces,logs & Deep Learning & Graphical neural model & NLP  & Knowledge Graph \\  
AutoTSG \cite{shetty2022autotsg} & Logs & Machine learning & Meta-learning algorithms  & - & Non-graph \\      
    \bottomrule  
  \end{tabular}  
\end{table*}
\subsubsection{Log Classification}
Log classification represents a principal method for root cause analysis in the domain of system log analysis, distinct from correlation-based approaches. It involves categorizing log entries into specific classes or types, often with a focus on differentiating normal system behavior from anomalous events that may indicate issues or failures. This classification can be based on various techniques, SAT-solver derives from Artificial Intelligence theories of action and diagnosis to identify the most relevant log entries for the given query \cite{Zawawy2010log}. \cite{gupta2015pariket, xu2017logdc, debnath2018loglens, 2019DeCaf, behera2023root} introduce machine learning approaches in log classification RCA. Pariket converts the event log into a sequential dataset with window-based and Markovian techniques and employ decision tree classifiers to extract rules elucidating the root causes \cite{gupta2015pariket}. DeCaf utilizes random forest to mine predicates from the logs that correlate with regressions \cite{2019DeCaf}. Logistic regression, naïve Bayes, decision trees, and random forests are introduced for comprehensive issue categorization RCA Bot \cite{behera2023root}. The extraction of relevant features or patterns and the application of clustering techniques are integral to the functionalities LogDC, LogLens, and HWLogAnalysis \cite{suriadi2013root, xu2017logdc, debnath2018loglens, tan2017hwlog}. These methods enable the systematic organization of log data by identifying and grouping similar entries, which facilitates the detection of anomalies and the subsequent root cause analysis. LADRA \cite{wanga2018ladra} employs the General Regression Neural Network (GRNN) to pinpoint the root causes of abnormal tasks. The likelihood of reported root causes is then presented to users based on weighted factors determined by the GRNN. Other methods seek failures similar logs or events correlated with failures, Markov decision model and Spectrum algorithms are used to pinpoint suspicious root cause \cite{ikeuchi2018root, rosenberg2020spectrum}.

Knowledge mining is a sophisticated technique used for extracting patterns, correlations, and causal relationships in log RCA. \cite{ding2014mining} employs natural language processing to assess similarities between new and historical issues logged. LogFaultFlagger \cite{amar2019mining} adopts a modified term frequency-inverse document frequency (TF-IDF) to identify the most relevant log lines related to past product failures and develops a specialized version of KNN to identify the faults. Log-Cluster \cite{lin2016log} constructs an initial knowledge base through clustering log sequences align with the knowledge base. Similarly, \cite{li2022abc} builds Knowledge Model (KM) to pinpoint root causes. Distinct from conventional methods that automatically detect root causes from given logs, they identify root cause metric by robust correlation analysis with data augmentation. 

Unlike previous works in incident management, which concentrate on complexities of incident triage, the construction of knowledge graphs leverages the graphical structure to collectively infer the root cause. By named-entity recognition (NER), relations entities mining from incidents, SoftNER \cite{shetty2022softner} facilitates the automatic construction of knowledge graphs. Incorporating an attention mechanism that takes advantage of both semantic context and data types to extract named entities, SoftNER performs root cause localization. Similarly, \cite{saha2022mining} builds a global Causal Knowledge Graph(CKG) to locate underlying the incidents.
\subsection{multi-Modal based Root Cause Analysis Techniques}
In the realm of complex software systems, diagnosing performance issues and identifying their underlying causes can be an arduous task. Methods relying on a single data source, such as metrics, traces, or logs, are insufficient, which may provide a narrow perspective and hinder a comprehensive understanding of the system’s behavior \cite{lee2023eadro, yu2023nezha}. This insufficiency is generated by each microservice individually at a local level, which violates the close dependency among microservices. To overcome this limitation, different with previous data-driven RCA methods relying on data from a single modality, researchers have explored multi-model root cause analysis. This approach integrates heterogeneous data sources, including metrics, traces, and logs, to offer a holistic and multi-faceted view of the system’s performance. 

In heterogeneous data sources, \cite{hou2021diagnosing, zheng2024multi} perform RCA based on the time series of events. Pdiagnose \cite{hou2021diagnosing} captures suspicious microservices and corresponding logs for logs and traces, and determines root causes through voting abnormal time series \cite{hou2021diagnosing}. MULAN \cite{zheng2024multi} leverages a log-tailored language model to facilitate log representation learning, converting raw log sequences into time-series data uses a contrastive learning-based approach to extract modality- invariant and modality-specific representations within a shared latent space, and introduces a novel key performance indicator-aware attention mechanism for assessing modality reliability and co-learning a final causal graph. MULAN employs random walk with restart to simulate system fault propagation and identify potential root causes.

Multi-model data source increases the complexity of RCA. By constructing O\&M knowledge graph or Causal Knowledge Graph (CKG), knowledge from multiple data sources will be integrated for further RCA \cite{qiu2020causality, liu2022microcbr, saha2022mining}. Optimized PC Algorithm based on the Knowledge Graph is used to construct a directed acyclic graph (DAG) Causality Graph, and breadth-first search (BFS) is employed to search for the root cause \cite{qiu2020causality}. From a knowledge base MicroCBR constructs spatial-temporal knowledge graph, and troubleshoots through case-based hierarchical reasoning \cite{liu2022microcbr}. SoTA neural NLP techniques are used to extract targeted knowledge from past incidents and they build a CKG, which is used to collectively infer the root cause through link prediction and graphical neural model methods \cite{saha2022mining}.

To make full use of heterogeneous multi-modal data, researchers capture dependencies from KPIs, topology, and logs. Both \cite{yu2023nezha, wang2021groot} capture dependencies from events, GROOT \cite{wang2021groot} collects all supported events for each service and constructs the causal links between events. Personalized PC algorithm and Granger Causality are also used for graph construction \cite{zhang2021cloudrca, white2021mmrca}. In CloudRCA, a Hierarchical Bayesian Network (KHBN) model, informed by knowledge, aids in inferring root causes on the constructed graph \cite{zhang2021cloudrca}. After heterogeneous multi-modal data are transferred into an (events) dependency graph, Nezha compares event patterns in the fault-free phase with those in the fault-affected phase to localize root causes in an interpretable manner \cite{yu2023nezha}. DiagFusion employs a graph neural network to identify the root cause instance and ascertain the failure type within the dependency graph \cite{zhang2023robust}. Eadro ranks the microservices based on their likelihood of being the source of the problem \cite{lee2023eadro}.
\setlength{\tabcolsep}{2pt}
\begin{table*}  
  \caption{Investigated root cause analysis approaches of multi-Model}  
  \label{tab:multi-Model}  
  \begin{tabular}{cp{2.5cm}cp{3cm}p{2.5cm}c} 
    \toprule  
    Method & Data source & RCA category & RCA method & Graph construction & Graph category \\  
    \midrule  
    Pdiagnose \cite{hou2021diagnosing} & Metrics, traces, logs & Statistical & Vote-based & - & Non-graph \\  
- \cite{qiu2020causality} & Metrics, traces, logs & Graph Theory & BFS & Optimized PC algorithm based on Knowledge Graph & Knowledge graph \\  
MicroCBR \cite{liu2022microcbr} & Metrics, logs, traces, commands & Statistical & Similarity comparison & DBSCAN clustering & Knowledge graph \\  
- \cite{saha2022mining} & Traces, logs & Deep Learning & Graphical neural model & NLP & Knowledge Graph \\  
CloudRCA \cite{zhang2021cloudrca} & KPIs, logs, traces & Machine learning & Bayesian network & PC algorithm & Dependency graph \\  
MMRCA \cite{white2021mmrca} & Metrics, traces, topology, configurations & Statistical & Personalized Granger Causality & Topology reduction & Dependency graph \\  
Eadro \cite{lee2023eadro} & Traces, logs, KPIs & Statistical & Culprit's probabilities similarity & Invocation relationship & Dependency graph \\  
DiagFusion \cite{zhang2023robust} & Metrics, traces, logs & Deep learning & GNN & Invocation relationship & Dependency graph \\  
Nezha \cite{yu2023nezha} & Metrics, traces, logs & Statistical & Pattern Miner and ranker & Personalized method & Dependency graph \\  
GROOT \cite{wang2021groot} & Metrics, logs, events & Graph Theory & Variant PageRank & Tracing and log analysis & Causality graph \\  
MULAN \cite{zheng2024multi} & Metrics, logs & Machine learning & Random walk & Personalized method & Causality graph \\  
    \bottomrule  
  \end{tabular}  
\end{table*}
\subsection{LLMs' Role in Enhancing RCA}
The recent emergence and remarkable performance of large language models (LLMs) in handling intricate tasks indicate a potential pathway for improving root cause analysis. Cloud providers and researchers have employed advanced AI-based solutions LLMs to aid SREs in identifying the root causes of cloud incidents. \cite{ahmed2023recommending} suggests fine-tuning large language models (LLMs) using domain-specific datasets, aiming to generate potential root causes of an incident solely based on the title and summary information available during the incident's creation. RCACopilot \cite{chen2023empowering} expands upon this work and add retrieval augmentation and diagnostic collection tools to the LLM-based root cause analysis pipeline. They design custom workflows for different types of incidents that trigger data collection procedures, which are then aggregated to predict a root cause category for the incident and help SREs with root cause analysis. RCACopilot relies on predefined handlers that must be engineered by hand. Considering the black-box nature of many LLM based methods, \cite{zhang2023pace} advocates fine-tuning or temperature-scaling-based approaches are inapplicable. \cite{zhang2023pace} introduces an innovative confidence estimation framework leveraging retrieval-augmented large language models. It assesses confidence in predicted root causes through two scoring phases, evaluating the model's "grounded-ness" in reference data and rating predictions based on historical references, followed by an optimization step to combine these scores. Besides, ADARMA \cite{sarda2023adarma} employs LLM based RCA for probabilistic reasoning, historical data, and correlation patterns to identify the likely root cause. \cite{sarda2023leveraging} uses semantic, lexical, and correctness metrics in zero-shot and few-shot scenarios.
In-context examples can serve as an alternative to fine-tuning for domain adaptation, but for agent-based RCA, crafting entire reasoning trajectories can be challenging \cite{roy2024exploring}. This is because agents require sophisticated prompting and typically also require fine-tuning \cite{yao2022react}. \cite{wang2023rcagent, chen2024automatic, roy2024exploring} explore LLM agents for root cause analysis. They all solve the problem of dynamic data collection based on ReAct agent \cite{yao2022react} to overcome the limitations of dynamically collect additional diagnostic information such as incident related metrics, logs or traces.
\section{Evaluation of RCA}
\subsection{Effectiveness and Efficiency}
When evaluating root cause analysis methods, effectiveness and efficiency serve as the primary criteria. MTTD (Mean Time to Diagnose) and MTTR play crucial roles in assessing the efficiency of these methods.
\setlength{\tabcolsep}{2pt}
\begin{table*}[h]
  \caption{Key Metrics of RCA Evaluation}  
  \label{tab:evaluation}  
  \begin{tabular}{cc} 
    \toprule  
    General metrics & Metrics  \\  
    \midrule  
Effectiveness & Top $n$ Accuracy, MAR, Precision, Recall, F1  \\  
Efficiency & Training time, Localization time \\  
    \bottomrule  
  \end{tabular}  
\end{table*}

(1) \textbf{Effectiveness} refers to the ability of the RCA system to accurately identify the true root causes of problems, which is prerequisite for interpretability. To assess the efficacy of associated methodologies, several key performance indicators are utilized: 

— Top $n$ Accuracy (A@K) denotes the likelihood that the correct outcome is included within the top K (K=1, 2, 3, …) suggestions provided by the fault localization technique. 

— Mean Average Rank (MAR, shown in Equation ~\ref{eq:MAR}) calculates the average position at which the true root causes are ranked across all instances of failure. This metric reflects the average number of recommendations that need to be reviewed in order to accurately diagnose a failure. Mean first rank (MFR) refers to the mean of the first root cause microservice rank in each fault. 
\begin{equation}
 \text{MAR} = \frac{1}{N} \sum_{i=1}^{N} \text{rank}_i \label{eq:MAR}
\end{equation}

— Besides, metrics like precision, recall, and F1 score are also often used to measure RCA algorithm’s effectiveness. Precision (Equation ~\ref{eq:Precision}) is the ratio of the number of correctly identified root causes to the total number of potential root causes proposed by the RCA system. This metric measures the proportion of true positive results (correctly identified root causes) in all positive results identified by the RCA method. It reflects the degree to which the system generates false positives in identifying root causes. TP represents the number of correctly identified root causes, and FP represents the number of incorrectly identified non-root causes. 
\begin{equation}
\text{Precision} = \nicefrac{\text{TP}}{(\text{TP} + \text{FP})} \tag{2}\label{eq:Precision}
\end{equation}

— Recall (Equation ~\ref{eq:Recall}) measures the proportion of true positive results in all actual positive cases (all actual root causes). It reflects the ability of the RCA system to identify all relevant root causes. A high recall rate indicates that the system is capable of capturing the majority of the actual root causes. FN represents the number of actual root causes that the system failed to identify. 
\begin{equation}
\text{Recall} = \nicefrac{\text{TP}}{(\text{TP} + \text{FN})} \tag{3}\label{eq:Recall}
\end{equation}

— The F1 (Equation ~\ref{eq:F1}) score is the harmonic mean of precision and recall, providing a single metric that takes into account both false positives and false negatives.
\begin{equation}
\text{F1} = 2 \times \left( \nicefrac{\text{Precision} \times \text{Recall}}{\text{Precision} + \text{Recall}} \right) \tag{4}\label{eq:F1}
\end{equation}

(2) \textbf{Efficiency} is a crucial metric that reflects the time and resources required by the online system to identify root causes. Efficiency typically encompasses two aspects for model-based approaches, training time and localization time. 

— Training time refers to the amount of time needed for an RCA model to learn and build a fault diagnosis model from training data. The formula for calculation is as follows: $T_{train} = T_{\text{end time of training}} - T_{\text{start time of training}}$, $T_{train}$ represents the training time.

— Localization time refers to the time it takes for an RCA system to output the root causes after detecting anomaly.  $T_{localize}$ can be formulated as follows: $T_{localize} = T_{\text{end time of localization}} - T_{\text{start time of localization}} $

In practical applications, especially in the RCA of large language models, the assessment of Efficiency may also include metrics of resource consumption such as GPU, memory required, processor utilization, and energy consumption to fully evaluate the system's efficiency.  For RCA systems that need real-time or near-real-time processing, localization time is particularly critical for MTTR, while training time is also not to be overlooked for continuously evolving systems. Thus, when assessing the efficiency of RCA methods, it is necessary to consider both time and resources dimensions comprehensively.
\subsection{Interpretability and Generalizability}
The interpretability of RCA refers to the ability of humans to understand and explain the decision-making process underlying the algorithm's outputs. In the context of complex system failures, where RCA is instrumental in identifying the root causes of incidents, interpretability is not just a desirable feature but often a necessary one. Engineers require clear and transparent reasoning to trust and effectively utilize the results of automated RCA methods. Explanations, given as potential failure cascades originating from an identified root cause, would enable intervention not only on the service that initially failed but also on those that failed in a cascading effect \cite{soldani2022anomaly}. In recent years, as advancements in algorithmic efficiency and accuracy have been achieved, an increasing number of researchers \cite{lin2020fast, li2022actionable, chen2023balance, yu2023nezha} have emphasized the importance of interpretability, with some even developing algorithms with interpretability as a primary objective. 

Interpretability evaluation involves examining how well the algorithm's logic, reasoning steps, and conclusions can be understood and followed by human analysts. \cite{lin2020fast} involves two filters for refining the findings in a production dataset: Filter 1 discards an item-set if there's a subset with similar or higher lift, indicating that shorter rules are preferred as they are considered more interesting. Filter 2 removes an item-set if there exists a superset with higher lift and similar support, suggesting that longer item-sets are kept only if they offer significantly more information. These techniques aim to balance the trade-off between the length of the rules (number of items) and their statistical significance (lift and support), enhancing the interpretability of the results. \cite{li2022actionable} involves a global interpretation approach using decision trees to provide understandable rules for identifying network-related issues, as well as a local interpretation technique that finds similar past failures to help engineers understand and address current network failures. \cite{yu2023nezha} designs an interpretable fine-grained framework to address the interpretability of RCA. 

Generalizability refers to the model's ability to process other dataset, that is, the model's capability to accurately predict or correctly handle data it has not seen before based on the knowledge learned from the training set. The generalizability of root cause analysis algorithms, beyond the algorithm itself, is affected by the architectural differences across various cloud environments, which also impacts the cross-platform compatibility and portability of AIOps algorithms. AIOps is not rented out as a capability in public clouds like other cloud products, largely due to the significant differences in underlying architectures, which in turn places higher demands on the generalizability of generic root cause analysis methods. Additionally, in a microservices environment, services and applications may continuously change and update. Therefore, a root cause analysis method with strong generalizability can maintain its performance across different environments, conditions, or datasets, which is particularly important for microservices architectures. \cite{li2022actionable} evaluates the generalizability by comparing the performance of previously seen (the ground truths are faulty in some historical failures) and unseen failures.

\section{Challenges \& Future Trends }
\subsection{Challenges}
Root Cause Analysis is a crucial methodology in identifying and rectifying faults that can occur within complex microservice environments. With the increasing adoption of microservices, effective RCA has become essential for maintaining the reliability, scalability, and performance of these distributed systems. Conventional RCA approaches, which depend on manual examination of data sources like logs and traces, tend to be arduous, susceptible to errors, and pose difficulties for engineers. By leveraging comprehensive data analysis from various sources and employing machine learning and other advanced AI techniques, RCA enables the automated identification of root causes, thereby accelerating resolution processes and reducing MTTR. Advanced RCA diagnostic methodologies are essential for both improving immediate response to faults and long-term system resilience.

Challenge in data control. RCA methods are full of challenges in data collection, quality and completeness. Traces and System logs can be voluminous and complex, containing vast amounts of information, which often contain noise and irrelevant information that can obfuscate the analysis. Distinguishing meaningful patterns from random fluctuations or irrelevant data is crucial for accurate RCA. Logs often contain limited contextual information, making it difficult to interpret and correlate log events with specific system components and behaviors. For graph-based methods, building a relational graph requires high-quality data with sufficient granularity. Incomplete, noisy, or biased data can lead to incorrect relationships or paths in RCA process. Graph construction can introduce ambiguities and identifying the actual root cause based on this graph can still be ambiguous. Methods that address this ambiguity determining the directionality and strength of relationships in the constructed graph requires careful consideration and validation. Besides, systems are constantly evolving with new updating, configurations, and interactions. Especially those based on microservices, can have dynamic topologies that change frequently. Keeping the relational graph up-to-date with the dynamic evolving system architectures, configurations, and operational changes is challenging. As the number of system components and interactions grows, the size of the graph can become unwieldy, leading to scalability issues in both graph construction and analysis.  

In view of the complexity of service relationship and the integrity of data, the RCA method of single dimension may be insufficient in practice. By integrating diverse data sources, including key performance indicators (KPIs), logs, traces, network topologies, and configurations, multi-modal RCA enables a comprehensive comprehension of system behaviors and dependencies. However, amalgamating heterogeneous data types into a cohesive framework poses a substantial challenge, primarily due to disparities in data formats, granularity, and temporal attributes. 

Challenge in RCA. For most data driven methods, effectively analyzing and correlating data to identify root causes remains a significant challenge. Graph-based RCA requires sophisticated algorithms to infer causality incorporating expert knowledge and automated approach. Distinguishing relationships between correlation and causation, and distinguishing causal relationships from temporal relationships are both inherently crucial in causal inference. In practice, traces may be incomplete or contain inaccuracies. These limitations can impact the accuracy and reliability of trace-based RCA. Effectively filtering and analyzing relevant logs to identify root causes remains a significant challenge. Incorporating additional context from system topology, configuration, and metadata can enhance RCA. Many failures involve complex causal chains and interactions between multiple components or services. Capturing these intricate failure patterns and identifying root causes accurately remains a challenge. Besides, systems evolve over time, with new components, dependencies, and behaviors emerging. RCA methods need to be able to adapt to system changes and identify new root causes effectively.

Despite the growing prevalence of AI-driven assistants in the root cause analysis process, the effectiveness in assisting on-call engineers is constrained by low accuracy due to the intrinsic difficulty of the task, a propensity for LLM-based approaches to hallucinate, and difficulties in distinguishing these well-disguised hallucinations. LLM agents are critical in collecting multi-modal data sources. LLM based finetuning requires various modal of complex unstructured or semi-structured data, uncompleted data might miss useful signals to reach conclusive diagnosis details. Finetuning is costly and may require a huge volume of training samples. Additionally, finetuning models are prone to generate more hallucinated results over time as its continuously updating a finetuned GPT model with evolving nature and scope of incidents. 

\subsection{Future Trends}
To meet the demands of modern systems and enable faster identification and resolution of breakdown, developing methods for real-time analysis will enable faster response times to incidents, minimizing MTTR. While metrics, traces, or logs provide valuable insights into root causes, sole data source RCA may not always be sufficient on their own. Causal inference should constitute a comprehensive end-to-end system that encompasses both the system infrastructure and the complexity of requests, in order to address the majority of fault scenarios. Future research should explore the integration with multi-scopes. For example, trace-based RCA can be enhanced by integrating it with other RCA methods, such as metric-based RCA and log analysis. Incorporating additional contextual information from system topology, configuration, and metadata into RCA techniques can provide a more comprehensive understanding of system behavior and failure patterns. 

Besides, fault understanding, incremental learning, and scalable analysis should offer a more holistic analysis of system failures. Advancements in AI and automation are anticipated to yield tools capable of verifying the dependability of constructed relational graphs, minimizing spurious correlations, enabling the automated and incremental construction of real-time graphs that adaptively evolve with system changes over time, thereby diminishing the necessity for manual intervention. From a system-level fault localization standpoint, multi-modal RCA embodies the prevailing trend in RCA. The advancement of unified models capable of handling multiple data types will prove increasingly significant in streamlining the analysis process and offering a comprehensive system perspective. The combined use of multiple techniques can provide a more holistic view of the system and improve the accuracy and effectiveness of RCA. 

Additionally, future research should focus on developing techniques that can provide detailed explanations and insights into the root causes identified, building user trust and enabling effective issue resolution. For example, explainable AI make the RCA process more transparent and understandable, allowing users to comprehend the logic behind the identified root causes. Advancements in AIOps have introduced new trends in RCA for microservices, large language model’s operations involve the use of powerful computational resources and algorithms by incorporating multidimensional dataset and end-to-end problem localization. These trends represent a shift towards more automated data-driven approaches to RCA.
\begin{acks}
This work is supported by the Natural Science Foundation of China (Grant No. U21A20488). We thank the Big Data Computing Center of Southeast University for providing the facility support on the numerical calculations in this paper.
\end{acks}
\bibliographystyle{ACM-Reference-Format}
\bibliography{sample-base}


\begin{thebibliography}{134}


\ifx \showCODEN    \undefined \def \showCODEN     #1{\unskip}     \fi
\ifx \showDOI      \undefined \def \showDOI       #1{#1}\fi
\ifx \showISBNx    \undefined \def \showISBNx     #1{\unskip}     \fi
\ifx \showISBNxiii \undefined \def \showISBNxiii  #1{\unskip}     \fi
\ifx \showISSN     \undefined \def \showISSN      #1{\unskip}     \fi
\ifx \showLCCN     \undefined \def \showLCCN      #1{\unskip}     \fi
\ifx \shownote     \undefined \def \shownote      #1{#1}          \fi
\ifx \showarticletitle \undefined \def \showarticletitle #1{#1}   \fi
\ifx \showURL      \undefined \def \showURL       {\relax}        \fi
\providecommand\bibfield[2]{#2}
\providecommand\bibinfo[2]{#2}
\providecommand\natexlab[1]{#1}
\providecommand\showeprint[2][]{arXiv:#2}

\bibitem[Ahmed et~al\mbox{.}(2023)]%
        {ahmed2023recommending}
\bibfield{author}{\bibinfo{person}{Toufique Ahmed}, \bibinfo{person}{Supriyo
  Ghosh}, \bibinfo{person}{Chetan Bansal}, \bibinfo{person}{Thomas Zimmermann},
  \bibinfo{person}{Xuchao Zhang}, {and} \bibinfo{person}{Saravan Rajmohan}.}
  \bibinfo{year}{2023}\natexlab{}.
\newblock \showarticletitle{Recommending Root-Cause and Mitigation Steps for
  Cloud Incidents using Large Language Models}.
\newblock \bibinfo{journal}{\emph{arXiv preprint arXiv:2301.03797}}
  (\bibinfo{year}{2023}).
\newblock


\bibitem[Aliyun(2022)]%
        {Aliyun20221218}
\bibfield{author}{\bibinfo{person}{Aliyun}.} \bibinfo{year}{2022}\natexlab{}.
\newblock \bibinfo{booktitle}{\emph{{Service Outage in Zone C of the China
  (Hong Kong) Region}}}.
\newblock


\bibitem[Aliyun(2023)]%
        {Aliyun20231127}
\bibfield{author}{\bibinfo{person}{Aliyun}.} \bibinfo{year}{2023}\natexlab{}.
\newblock \bibinfo{booktitle}{\emph{{Aliyun Cloud Product Console Service
  Anomaly}}}.
\newblock


\bibitem[Amar and Rigby(2019)]%
        {amar2019mining}
\bibfield{author}{\bibinfo{person}{Anunay Amar} {and} \bibinfo{person}{Peter~C
  Rigby}.} \bibinfo{year}{2019}\natexlab{}.
\newblock \showarticletitle{Mining historical test logs to predict bugs and
  localize faults in the test logs}. In \bibinfo{booktitle}{\emph{2019 IEEE/ACM
  41st International Conference on Software Engineering (ICSE)}}. IEEE,
  \bibinfo{pages}{140--151}.
\newblock


\bibitem[Amazon(2021)]%
        {Amazon2021}
\bibfield{author}{\bibinfo{person}{Amazon}.} \bibinfo{year}{2021}\natexlab{}.
\newblock \bibinfo{booktitle}{\emph{{Summary of the AWS Service Event in the
  Northern Virginia (US-EAST-1) Region}}}.
\newblock


\bibitem[Attariyan et~al\mbox{.}(2012)]%
        {attariyan2012x}
\bibfield{author}{\bibinfo{person}{Mona Attariyan}, \bibinfo{person}{Michael
  Chow}, {and} \bibinfo{person}{Jason Flinn}.} \bibinfo{year}{2012}\natexlab{}.
\newblock \showarticletitle{X-ray: automating $\{$Root-Cause$\}$ diagnosis of
  performance anomalies in production software}. In
  \bibinfo{booktitle}{\emph{10th USENIX Symposium on Operating Systems Design
  and Implementation (OSDI 12)}}. \bibinfo{pages}{307--320}.
\newblock


\bibitem[baidu baike(2023)]%
        {Didi2023}
\bibfield{author}{\bibinfo{person}{baidu baike}.}
  \bibinfo{year}{2023}\natexlab{}.
\newblock \bibinfo{booktitle}{\emph{{11·27 Didi Outage Incident}}}.
\newblock


\bibitem[Bansal et~al\mbox{.}(2019)]%
        {2019DeCaf}
\bibfield{author}{\bibinfo{person}{Chetan Bansal},
  \bibinfo{person}{Sundararajan Renganathan}, \bibinfo{person}{Ashima Asudani},
  \bibinfo{person}{Olivier Midy}, {and} \bibinfo{person}{Mathru Janakiraman}.}
  \bibinfo{year}{2019}\natexlab{}.
\newblock \showarticletitle{DeCaf: Diagnosing and Triaging Performance Issues
  in Large-Scale Cloud Services}.
\newblock  (\bibinfo{year}{2019}).
\newblock


\bibitem[Behera et~al\mbox{.}(2023)]%
        {behera2023trace}
\bibfield{author}{\bibinfo{person}{Anukampa Behera},
  \bibinfo{person}{Chhabi~Rani Panigrahi}, \bibinfo{person}{Sitesh Behera},
  \bibinfo{person}{Rohit Patel}, {and} \bibinfo{person}{Sourav Bera}.}
  \bibinfo{year}{2023}\natexlab{}.
\newblock \showarticletitle{trACE-Anomaly Correlation Engine for Tracing the
  Root Cause on Cloud Based Microservice Architecture}.
\newblock \bibinfo{journal}{\emph{Computaci{\'o}n y Sistemas}}
  \bibinfo{volume}{27}, \bibinfo{number}{3} (\bibinfo{year}{2023}),
  \bibinfo{pages}{791--800}.
\newblock


\bibitem[Behera and Tripathi(2023)]%
        {behera2023root}
\bibfield{author}{\bibinfo{person}{Tapan Behera} {and} \bibinfo{person}{Kumud
  Tripathi}.} \bibinfo{year}{2023}\natexlab{}.
\newblock \showarticletitle{Root Cause Analysis Bot using Machine Learning
  Techniques}.
\newblock \bibinfo{journal}{\emph{Authorea Preprints}} (\bibinfo{year}{2023}).
\newblock


\bibitem[Bilibili(2021)]%
        {Bilibili2021}
\bibfield{author}{\bibinfo{person}{Bilibili}.} \bibinfo{year}{2021}\natexlab{}.
\newblock \bibinfo{booktitle}{\emph{{On July 13, 2021, we collapsed in this
  way.}}}
\newblock


\bibitem[Brand{\'o}n et~al\mbox{.}(2020)]%
        {brandon2020graph}
\bibfield{author}{\bibinfo{person}{{\'A}lvaro Brand{\'o}n},
  \bibinfo{person}{Marc Sol{\'e}}, \bibinfo{person}{Alberto Hu{\'e}lamo},
  \bibinfo{person}{David Solans}, \bibinfo{person}{Mar{\'\i}a~S P{\'e}rez},
  {and} \bibinfo{person}{Victor Munt{\'e}s-Mulero}.}
  \bibinfo{year}{2020}\natexlab{}.
\newblock \showarticletitle{Graph-based root cause analysis for
  service-oriented and microservice architectures}.
\newblock \bibinfo{journal}{\emph{Journal of Systems and Software}}
  \bibinfo{volume}{159} (\bibinfo{year}{2020}), \bibinfo{pages}{110432}.
\newblock


\bibitem[Cai et~al\mbox{.}(2019)]%
        {cai2019real}
\bibfield{author}{\bibinfo{person}{Zhengong Cai}, \bibinfo{person}{Wei Li},
  \bibinfo{person}{Wanyi Zhu}, \bibinfo{person}{Lu Liu}, {and}
  \bibinfo{person}{Bowei Yang}.} \bibinfo{year}{2019}\natexlab{}.
\newblock \showarticletitle{A real-time trace-level root-cause diagnosis system
  in alibaba datacenters}.
\newblock \bibinfo{journal}{\emph{IEEE Access}}  \bibinfo{volume}{7}
  (\bibinfo{year}{2019}), \bibinfo{pages}{142692--142702}.
\newblock


\bibitem[Chakraborty et~al\mbox{.}(2023)]%
        {chakraborty2023causil}
\bibfield{author}{\bibinfo{person}{Sarthak Chakraborty},
  \bibinfo{person}{Shaddy Garg}, \bibinfo{person}{Shubham Agarwal},
  \bibinfo{person}{Ayush Chauhan}, {and} \bibinfo{person}{Shiv~Kumar Saini}.}
  \bibinfo{year}{2023}\natexlab{}.
\newblock \showarticletitle{Causil: Causal graph for instance level
  microservice data}. In \bibinfo{booktitle}{\emph{Proceedings of the ACM Web
  Conference 2023}}. \bibinfo{pages}{2905--2915}.
\newblock


\bibitem[Chen et~al\mbox{.}(2023b)]%
        {chen2023balance}
\bibfield{author}{\bibinfo{person}{Chaoyu Chen}, \bibinfo{person}{Hang Yu},
  \bibinfo{person}{Zhichao Lei}, \bibinfo{person}{Jianguo Li},
  \bibinfo{person}{Shaokang Ren}, \bibinfo{person}{Tingkai Zhang},
  \bibinfo{person}{Silin Hu}, \bibinfo{person}{Jianchao Wang}, {and}
  \bibinfo{person}{Wenhui Shi}.} \bibinfo{year}{2023}\natexlab{b}.
\newblock \showarticletitle{BALANCE: Bayesian Linear Attribution for Root Cause
  Localization}.
\newblock \bibinfo{journal}{\emph{Proceedings of the ACM on Management of
  Data}} \bibinfo{volume}{1}, \bibinfo{number}{1} (\bibinfo{year}{2023}),
  \bibinfo{pages}{1--26}.
\newblock


\bibitem[Chen et~al\mbox{.}(2021)]%
        {chen2021trace}
\bibfield{author}{\bibinfo{person}{Hao Chen}, \bibinfo{person}{Kegang Wei},
  \bibinfo{person}{An Li}, \bibinfo{person}{Tao Wang}, {and}
  \bibinfo{person}{Wenbo Zhang}.} \bibinfo{year}{2021}\natexlab{}.
\newblock \showarticletitle{Trace-based intelligent fault diagnosis for
  microservices with deep learning}. In \bibinfo{booktitle}{\emph{2021 IEEE
  45th Annual Computers, Software, and Applications Conference (COMPSAC)}}.
  IEEE, \bibinfo{pages}{884--893}.
\newblock


\bibitem[Chen et~al\mbox{.}(2002)]%
        {chen2002pinpoint}
\bibfield{author}{\bibinfo{person}{Mike~Y Chen}, \bibinfo{person}{Emre
  Kiciman}, \bibinfo{person}{Eugene Fratkin}, \bibinfo{person}{Armando Fox},
  {and} \bibinfo{person}{Eric Brewer}.} \bibinfo{year}{2002}\natexlab{}.
\newblock \showarticletitle{Pinpoint: Problem determination in large, dynamic
  internet services}. In \bibinfo{booktitle}{\emph{Proceedings International
  Conference on Dependable Systems and Networks}}. IEEE,
  \bibinfo{pages}{595--604}.
\newblock


\bibitem[Chen et~al\mbox{.}(2014)]%
        {chen2014causeinfer}
\bibfield{author}{\bibinfo{person}{Pengfei Chen}, \bibinfo{person}{Yong Qi},
  \bibinfo{person}{Pengfei Zheng}, {and} \bibinfo{person}{Di Hou}.}
  \bibinfo{year}{2014}\natexlab{}.
\newblock \showarticletitle{Causeinfer: Automatic and distributed performance
  diagnosis with hierarchical causality graph in large distributed systems}. In
  \bibinfo{booktitle}{\emph{IEEE INFOCOM 2014-IEEE Conference on Computer
  Communications}}. IEEE, \bibinfo{pages}{1887--1895}.
\newblock


\bibitem[Chen et~al\mbox{.}(2023a)]%
        {chen2023empowering}
\bibfield{author}{\bibinfo{person}{Yinfang Chen}, \bibinfo{person}{Huaibing
  Xie}, \bibinfo{person}{Minghua Ma}, \bibinfo{person}{Yu Kang},
  \bibinfo{person}{Xin Gao}, \bibinfo{person}{Liu Shi}, \bibinfo{person}{Yunjie
  Cao}, \bibinfo{person}{Xuedong Gao}, \bibinfo{person}{Hao Fan},
  \bibinfo{person}{Ming Wen}, {et~al\mbox{.}}}
  \bibinfo{year}{2023}\natexlab{a}.
\newblock \showarticletitle{Empowering practical root cause analysis by large
  language models for cloud incidents}.
\newblock \bibinfo{journal}{\emph{arXiv preprint arXiv:2305.15778}}
  (\bibinfo{year}{2023}).
\newblock


\bibitem[Chen et~al\mbox{.}(2024)]%
        {chen2024automatic}
\bibfield{author}{\bibinfo{person}{Yinfang Chen}, \bibinfo{person}{Huaibing
  Xie}, \bibinfo{person}{Minghua Ma}, \bibinfo{person}{Yu Kang},
  \bibinfo{person}{Xin Gao}, \bibinfo{person}{Liu Shi}, \bibinfo{person}{Yunjie
  Cao}, \bibinfo{person}{Xuedong Gao}, \bibinfo{person}{Hao Fan},
  \bibinfo{person}{Ming Wen}, {et~al\mbox{.}}} \bibinfo{year}{2024}\natexlab{}.
\newblock \showarticletitle{Automatic Root Cause Analysis via Large Language
  Models for Cloud Incidents}.
\newblock  (\bibinfo{year}{2024}).
\newblock


\bibitem[Cheng et~al\mbox{.}(2023)]%
        {cheng2023ai}
\bibfield{author}{\bibinfo{person}{Qian Cheng}, \bibinfo{person}{Doyen Sahoo},
  \bibinfo{person}{Amrita Saha}, \bibinfo{person}{Wenzhuo Yang},
  \bibinfo{person}{Chenghao Liu}, \bibinfo{person}{Gerald Woo},
  \bibinfo{person}{Manpreet Singh}, \bibinfo{person}{Silvio Saverese}, {and}
  \bibinfo{person}{Steven~CH Hoi}.} \bibinfo{year}{2023}\natexlab{}.
\newblock \showarticletitle{Ai for it operations (aiops) on cloud platforms:
  Reviews, opportunities and challenges}.
\newblock \bibinfo{journal}{\emph{arXiv preprint arXiv:2304.04661}}
  (\bibinfo{year}{2023}).
\newblock


\bibitem[Chuah et~al\mbox{.}(2010)]%
        {chuah2010diagnosing}
\bibfield{author}{\bibinfo{person}{Edward Chuah}, \bibinfo{person}{Shyh-hao
  Kuo}, \bibinfo{person}{Paul Hiew}, \bibinfo{person}{William-Chandra Tjhi},
  \bibinfo{person}{Gary Lee}, \bibinfo{person}{John Hammond},
  \bibinfo{person}{Marek~T Michalewicz}, \bibinfo{person}{Terence Hung}, {and}
  \bibinfo{person}{James~C Browne}.} \bibinfo{year}{2010}\natexlab{}.
\newblock \showarticletitle{Diagnosing the root-causes of failures from cluster
  log files}. In \bibinfo{booktitle}{\emph{2010 International Conference on
  High Performance Computing}}. IEEE, \bibinfo{pages}{1--10}.
\newblock


\bibitem[Cinque et~al\mbox{.}(2012)]%
        {cinque2012event}
\bibfield{author}{\bibinfo{person}{Marcello Cinque}, \bibinfo{person}{Domenico
  Cotroneo}, {and} \bibinfo{person}{Antonio Pecchia}.}
  \bibinfo{year}{2012}\natexlab{}.
\newblock \showarticletitle{Event logs for the analysis of software failures: A
  rule-based approach}.
\newblock \bibinfo{journal}{\emph{IEEE Transactions on Software Engineering}}
  \bibinfo{volume}{39}, \bibinfo{number}{6} (\bibinfo{year}{2012}),
  \bibinfo{pages}{806--821}.
\newblock


\bibitem[Cloud(2024)]%
        {TencentCloud2024}
\bibfield{author}{\bibinfo{person}{Tencent Cloud}.}
  \bibinfo{year}{2024}\natexlab{}.
\newblock \bibinfo{booktitle}{\emph{{Tencent Cloud Situation Report of the
  April 8th Outage}}}.
\newblock


\bibitem[CNN(2023)]%
        {Twitter2023}
\bibfield{author}{\bibinfo{person}{CNN}.} \bibinfo{year}{2023}\natexlab{}.
\newblock \bibinfo{booktitle}{\emph{{Twitter hit with one of the biggest
  outages since Elon Musk took over.}}}
\newblock


\bibitem[Debnath et~al\mbox{.}(2018)]%
        {debnath2018loglens}
\bibfield{author}{\bibinfo{person}{Biplob Debnath}, \bibinfo{person}{Mohiuddin
  Solaimani}, \bibinfo{person}{Muhammad Ali~Gulzar Gulzar},
  \bibinfo{person}{Nipun Arora}, \bibinfo{person}{Cristian Lumezanu},
  \bibinfo{person}{Jianwu Xu}, \bibinfo{person}{Bo Zong}, \bibinfo{person}{Hui
  Zhang}, \bibinfo{person}{Guofei Jiang}, {and} \bibinfo{person}{Latifur
  Khan}.} \bibinfo{year}{2018}\natexlab{}.
\newblock \showarticletitle{Loglens: A real-time log analysis system}. In
  \bibinfo{booktitle}{\emph{2018 IEEE 38th international conference on
  distributed computing systems (ICDCS)}}. IEEE, \bibinfo{pages}{1052--1062}.
\newblock


\bibitem[Ding et~al\mbox{.}(2014)]%
        {ding2014mining}
\bibfield{author}{\bibinfo{person}{Rui Ding}, \bibinfo{person}{Qiang Fu},
  \bibinfo{person}{Jian~Guang Lou}, \bibinfo{person}{Qingwei Lin},
  \bibinfo{person}{Dongmei Zhang}, {and} \bibinfo{person}{Tao Xie}.}
  \bibinfo{year}{2014}\natexlab{}.
\newblock \showarticletitle{Mining historical issue repositories to heal
  large-scale online service systems}. In \bibinfo{booktitle}{\emph{2014 44th
  Annual IEEE/IFIP International Conference on Dependable Systems and
  Networks}}. IEEE, \bibinfo{pages}{311--322}.
\newblock


\bibitem[ENOW(2023)]%
        {Microsoft20230605}
\bibfield{author}{\bibinfo{person}{ENOW}.} \bibinfo{year}{2023}\natexlab{}.
\newblock \bibinfo{booktitle}{\emph{{Microsoft Outlook Suffers Major Outage}}}.
\newblock


\bibitem[Fu et~al\mbox{.}(2014)]%
        {fu2014digging}
\bibfield{author}{\bibinfo{person}{Xiaoyu Fu}, \bibinfo{person}{Rui Ren},
  \bibinfo{person}{Sally~A McKee}, \bibinfo{person}{Jianfeng Zhan}, {and}
  \bibinfo{person}{Ninghui Sun}.} \bibinfo{year}{2014}\natexlab{}.
\newblock \showarticletitle{Digging deeper into cluster system logs for failure
  prediction and root cause diagnosis}. In \bibinfo{booktitle}{\emph{2014 IEEE
  International Conference on Cluster Computing (CLUSTER)}}. IEEE,
  \bibinfo{pages}{103--112}.
\newblock


\bibitem[{FudanSELab}(2024)]%
        {FudanSELab_train_ticket}
\bibfield{author}{\bibinfo{person}{{FudanSELab}}.}
  \bibinfo{year}{2024}\natexlab{}.
\newblock \bibinfo{title}{FudanSELab/train-ticket}.
\newblock
\newblock
\urldef\tempurl%
\url{https://github.com/FudanSELab/train-ticket}
\showURL{%
\tempurl}
\newblock
\shownote{Accessed: 2024-5-26}.


\bibitem[Gan et~al\mbox{.}(2021)]%
        {gan2021sage}
\bibfield{author}{\bibinfo{person}{Yu Gan}, \bibinfo{person}{Mingyu Liang},
  \bibinfo{person}{Sundar Dev}, \bibinfo{person}{David Lo}, {and}
  \bibinfo{person}{Christina Delimitrou}.} \bibinfo{year}{2021}\natexlab{}.
\newblock \showarticletitle{Sage: practical and scalable ML-driven performance
  debugging in microservices}. In \bibinfo{booktitle}{\emph{Proceedings of the
  26th ACM International Conference on Architectural Support for Programming
  Languages and Operating Systems}}. \bibinfo{pages}{135--151}.
\newblock


\bibitem[Gan et~al\mbox{.}(2023)]%
        {gan2023sleuth}
\bibfield{author}{\bibinfo{person}{Yu Gan}, \bibinfo{person}{Guiyang Liu},
  \bibinfo{person}{Xin Zhang}, \bibinfo{person}{Qi Zhou},
  \bibinfo{person}{Jiesheng Wu}, {and} \bibinfo{person}{Jiangwei Jiang}.}
  \bibinfo{year}{2023}\natexlab{}.
\newblock \showarticletitle{Sleuth: A Trace-Based Root Cause Analysis System
  for Large-Scale Microservices with Graph Neural Networks}. In
  \bibinfo{booktitle}{\emph{Proceedings of the 28th ACM International
  Conference on Architectural Support for Programming Languages and Operating
  Systems, Volume 4}}. \bibinfo{pages}{324--337}.
\newblock


\bibitem[Gertler(2017)]%
        {gertler2017fault}
\bibfield{author}{\bibinfo{person}{Janos Gertler}.}
  \bibinfo{year}{2017}\natexlab{}.
\newblock \bibinfo{booktitle}{\emph{Fault detection and diagnosis in
  engineering systems}}.
\newblock \bibinfo{publisher}{CRC press}.
\newblock


\bibitem[Guo et~al\mbox{.}(2020)]%
        {guo2020graph}
\bibfield{author}{\bibinfo{person}{Xiaofeng Guo}, \bibinfo{person}{Xin Peng},
  \bibinfo{person}{Hanzhang Wang}, \bibinfo{person}{Wanxue Li},
  \bibinfo{person}{Huai Jiang}, \bibinfo{person}{Dan Ding},
  \bibinfo{person}{Tao Xie}, {and} \bibinfo{person}{Liangfei Su}.}
  \bibinfo{year}{2020}\natexlab{}.
\newblock \showarticletitle{Graph-based trace analysis for microservice
  architecture understanding and problem diagnosis}. In
  \bibinfo{booktitle}{\emph{Proceedings of the 28th ACM Joint Meeting on
  European Software Engineering Conference and Symposium on the Foundations of
  Software Engineering}}. \bibinfo{pages}{1387--1397}.
\newblock


\bibitem[Gupta et~al\mbox{.}(2003)]%
        {2003Discovering}
\bibfield{author}{\bibinfo{person}{Manish Gupta}, \bibinfo{person}{Anindya
  Neogi}, \bibinfo{person}{Manoj~K. Agarwal}, {and} \bibinfo{person}{Gautam
  Kar}.} \bibinfo{year}{2003}\natexlab{}.
\newblock \showarticletitle{Discovering Dynamic Dependencies in Enterprise
  Environments for Problem Determination}. In \bibinfo{booktitle}{\emph{IEEE
  International Workshop on Self-managing Distributed Systems}}.
\newblock


\bibitem[Gupta et~al\mbox{.}(2015)]%
        {gupta2015pariket}
\bibfield{author}{\bibinfo{person}{Nisha Gupta}, \bibinfo{person}{Kritika
  Anand}, {and} \bibinfo{person}{Ashish Sureka}.}
  \bibinfo{year}{2015}\natexlab{}.
\newblock \showarticletitle{Pariket: Mining business process logs for root
  cause analysis of anomalous incidents}. In
  \bibinfo{booktitle}{\emph{Databases in Networked Information Systems: 10th
  International Workshop, DNIS 2015, Aizu-Wakamatsu, Japan, March 23-25, 2015.
  Proceedings 10}}. Springer, \bibinfo{pages}{244--263}.
\newblock


\bibitem[{helidon-sockshop}(2024)]%
        {helidon_sockshop_sockshop}
\bibfield{author}{\bibinfo{person}{{helidon-sockshop}}.}
  \bibinfo{year}{2024}\natexlab{}.
\newblock \bibinfo{title}{helidon-sockshop/sockshop}.
\newblock
\newblock
\urldef\tempurl%
\url{https://github.com/helidon-sockshop/sockshop}
\showURL{%
\tempurl}
\newblock
\shownote{Accessed: 2024-5-26}.


\bibitem[Hou et~al\mbox{.}(2021)]%
        {hou2021diagnosing}
\bibfield{author}{\bibinfo{person}{Chuanjia Hou}, \bibinfo{person}{Tong Jia},
  \bibinfo{person}{Yifan Wu}, \bibinfo{person}{Ying Li}, {and}
  \bibinfo{person}{Jing Han}.} \bibinfo{year}{2021}\natexlab{}.
\newblock \showarticletitle{Diagnosing performance issues in microservices with
  heterogeneous data source}. In \bibinfo{booktitle}{\emph{2021 IEEE Intl Conf
  on Parallel \& Distributed Processing with Applications, Big Data \& Cloud
  Computing, Sustainable Computing \& Communications, Social Computing \&
  Networking (ISPA/BDCloud/SocialCom/SustainCom)}}. IEEE,
  \bibinfo{pages}{493--500}.
\newblock


\bibitem[Ikeuchi et~al\mbox{.}(2018)]%
        {ikeuchi2018root}
\bibfield{author}{\bibinfo{person}{Hiroki Ikeuchi}, \bibinfo{person}{Akio
  Watanabe}, \bibinfo{person}{Takehiro Kawata}, {and} \bibinfo{person}{Ryoichi
  Kawahara}.} \bibinfo{year}{2018}\natexlab{}.
\newblock \showarticletitle{Root-cause diagnosis using logs generated by user
  actions}. In \bibinfo{booktitle}{\emph{2018 IEEE Global Communications
  Conference (GLOBECOM)}}. IEEE, \bibinfo{pages}{1--7}.
\newblock


\bibitem[Jia et~al\mbox{.}(2017)]%
        {jia2017logsed}
\bibfield{author}{\bibinfo{person}{Tong Jia}, \bibinfo{person}{Lin Yang},
  \bibinfo{person}{Pengfei Chen}, \bibinfo{person}{Ying Li},
  \bibinfo{person}{Fanjing Meng}, {and} \bibinfo{person}{Jingmin Xu}.}
  \bibinfo{year}{2017}\natexlab{}.
\newblock \showarticletitle{Logsed: Anomaly diagnosis through mining
  time-weighted control flow graph in logs}. In \bibinfo{booktitle}{\emph{2017
  IEEE 10th International Conference on Cloud Computing (CLOUD)}}. IEEE,
  \bibinfo{pages}{447--455}.
\newblock


\bibitem[Julisch(2003)]%
        {julisch2003clustering}
\bibfield{author}{\bibinfo{person}{Klaus Julisch}.}
  \bibinfo{year}{2003}\natexlab{}.
\newblock \showarticletitle{Clustering intrusion detection alarms to support
  root cause analysis}.
\newblock \bibinfo{journal}{\emph{ACM transactions on information and system
  security (TISSEC)}} \bibinfo{volume}{6}, \bibinfo{number}{4}
  (\bibinfo{year}{2003}), \bibinfo{pages}{443--471}.
\newblock


\bibitem[Kim et~al\mbox{.}(2013)]%
        {kim2013root}
\bibfield{author}{\bibinfo{person}{Myunghwan Kim}, \bibinfo{person}{Roshan
  Sumbaly}, {and} \bibinfo{person}{Sam Shah}.} \bibinfo{year}{2013}\natexlab{}.
\newblock \showarticletitle{Root cause detection in a service-oriented
  architecture}.
\newblock \bibinfo{journal}{\emph{ACM SIGMETRICS Performance Evaluation
  Review}} \bibinfo{volume}{41}, \bibinfo{number}{1} (\bibinfo{year}{2013}),
  \bibinfo{pages}{93--104}.
\newblock


\bibitem[Koyama and Kushida(2023)]%
        {koyama2023log}
\bibfield{author}{\bibinfo{person}{Tomoyuki Koyama} {and}
  \bibinfo{person}{Takayuki Kushida}.} \bibinfo{year}{2023}\natexlab{}.
\newblock \showarticletitle{Log message with JSON item count for root cause
  analysis in microservices}. In \bibinfo{booktitle}{\emph{2023 6th Conference
  on Cloud and Internet of Things (CIoT)}}. IEEE, \bibinfo{pages}{55--61}.
\newblock


\bibitem[Lee et~al\mbox{.}(2023)]%
        {lee2023eadro}
\bibfield{author}{\bibinfo{person}{Cheryl Lee}, \bibinfo{person}{Tianyi Yang},
  \bibinfo{person}{Zhuangbin Chen}, \bibinfo{person}{Yuxin Su}, {and}
  \bibinfo{person}{Michael~R Lyu}.} \bibinfo{year}{2023}\natexlab{}.
\newblock \showarticletitle{Eadro: An end-to-end troubleshooting framework for
  microservices on multi-source data}. In \bibinfo{booktitle}{\emph{2023
  IEEE/ACM 45th International Conference on Software Engineering (ICSE)}}.
  IEEE, \bibinfo{pages}{1750--1762}.
\newblock


\bibitem[Li et~al\mbox{.}(2022b)]%
        {li2022causal}
\bibfield{author}{\bibinfo{person}{Mingjie Li}, \bibinfo{person}{Zeyan Li},
  \bibinfo{person}{Kanglin Yin}, \bibinfo{person}{Xiaohui Nie},
  \bibinfo{person}{Wenchi Zhang}, \bibinfo{person}{Kaixin Sui}, {and}
  \bibinfo{person}{Dan Pei}.} \bibinfo{year}{2022}\natexlab{b}.
\newblock \showarticletitle{Causal inference-based root cause analysis for
  online service systems with intervention recognition}. In
  \bibinfo{booktitle}{\emph{Proceedings of the 28th ACM SIGKDD Conference on
  Knowledge Discovery and Data Mining}}. \bibinfo{pages}{3230--3240}.
\newblock


\bibitem[Li et~al\mbox{.}(2022c)]%
        {li2022mining}
\bibfield{author}{\bibinfo{person}{Mingjie Li}, \bibinfo{person}{Minghua Ma},
  \bibinfo{person}{Xiaohui Nie}, \bibinfo{person}{Kanglin Yin},
  \bibinfo{person}{Li Cao}, \bibinfo{person}{Xidao Wen},
  \bibinfo{person}{Zhiyun Yuan}, \bibinfo{person}{Duogang Wu},
  \bibinfo{person}{Guoying Li}, \bibinfo{person}{Wei Liu}, {et~al\mbox{.}}}
  \bibinfo{year}{2022}\natexlab{c}.
\newblock \showarticletitle{Mining Fluctuation Propagation Graph Among Time
  Series with Active Learning}. In \bibinfo{booktitle}{\emph{International
  Conference on Database and Expert Systems Applications}}. Springer,
  \bibinfo{pages}{220--233}.
\newblock


\bibitem[Li et~al\mbox{.}(2022a)]%
        {li2022abc}
\bibfield{author}{\bibinfo{person}{Xue Li}, \bibinfo{person}{Alan Bundy},
  \bibinfo{person}{Ruiqi Zhu}, \bibinfo{person}{Fangrong Wang},
  \bibinfo{person}{Stefano Mauceri}, \bibinfo{person}{Lei Xu}, {and}
  \bibinfo{person}{Jeff~Z Pan}.} \bibinfo{year}{2022}\natexlab{a}.
\newblock \showarticletitle{ABC in Root Cause Analysis: Discovering Missing
  Information and Repairing System Failures}. In
  \bibinfo{booktitle}{\emph{International Conference on Machine Learning,
  Optimization, and Data Science}}. Springer, \bibinfo{pages}{346--359}.
\newblock


\bibitem[Li et~al\mbox{.}(2021)]%
        {li2021practical}
\bibfield{author}{\bibinfo{person}{Zeyan Li}, \bibinfo{person}{Junjie Chen},
  \bibinfo{person}{Rui Jiao}, \bibinfo{person}{Nengwen Zhao},
  \bibinfo{person}{Zhijun Wang}, \bibinfo{person}{Shuwei Zhang},
  \bibinfo{person}{Yanjun Wu}, \bibinfo{person}{Long Jiang},
  \bibinfo{person}{Leiqin Yan}, \bibinfo{person}{Zikai Wang}, {et~al\mbox{.}}}
  \bibinfo{year}{2021}\natexlab{}.
\newblock \showarticletitle{Practical root cause localization for microservice
  systems via trace analysis}. In \bibinfo{booktitle}{\emph{2021 IEEE/ACM 29th
  International Symposium on Quality of Service (IWQOS)}}. IEEE,
  \bibinfo{pages}{1--10}.
\newblock


\bibitem[Li et~al\mbox{.}(2022d)]%
        {li2022actionable}
\bibfield{author}{\bibinfo{person}{Zeyan Li}, \bibinfo{person}{Nengwen Zhao},
  \bibinfo{person}{Mingjie Li}, \bibinfo{person}{Xianglin Lu},
  \bibinfo{person}{Lixin Wang}, \bibinfo{person}{Dongdong Chang},
  \bibinfo{person}{Xiaohui Nie}, \bibinfo{person}{Li Cao},
  \bibinfo{person}{Wenchi Zhang}, \bibinfo{person}{Kaixin Sui},
  {et~al\mbox{.}}} \bibinfo{year}{2022}\natexlab{d}.
\newblock \showarticletitle{Actionable and interpretable fault localization for
  recurring failures in online service systems}. In
  \bibinfo{booktitle}{\emph{Proceedings of the 30th ACM Joint European Software
  Engineering Conference and Symposium on the Foundations of Software
  Engineering}}. \bibinfo{pages}{996--1008}.
\newblock


\bibitem[Lin et~al\mbox{.}(2020)]%
        {lin2020fast}
\bibfield{author}{\bibinfo{person}{Fred Lin}, \bibinfo{person}{Keyur Muzumdar},
  \bibinfo{person}{Nikolay~Pavlovich Laptev}, \bibinfo{person}{Mihai-Valentin
  Curelea}, \bibinfo{person}{Seunghak Lee}, {and} \bibinfo{person}{Sriram
  Sankar}.} \bibinfo{year}{2020}\natexlab{}.
\newblock \showarticletitle{Fast dimensional analysis for root cause
  investigation in a large-scale service environment}.
\newblock \bibinfo{journal}{\emph{Proceedings of the ACM on Measurement and
  Analysis of Computing Systems}} \bibinfo{volume}{4}, \bibinfo{number}{2}
  (\bibinfo{year}{2020}), \bibinfo{pages}{1--23}.
\newblock


\bibitem[Lin et~al\mbox{.}(2018a)]%
        {lin2018microscope}
\bibfield{author}{\bibinfo{person}{JinJin Lin}, \bibinfo{person}{Pengfei Chen},
  {and} \bibinfo{person}{Zibin Zheng}.} \bibinfo{year}{2018}\natexlab{a}.
\newblock \showarticletitle{Microscope: Pinpoint performance issues with causal
  graphs in micro-service environments}. In
  \bibinfo{booktitle}{\emph{Service-Oriented Computing: 16th International
  Conference, ICSOC 2018, Hangzhou, China, November 12-15, 2018, Proceedings
  16}}. Springer, \bibinfo{pages}{3--20}.
\newblock


\bibitem[Lin et~al\mbox{.}(2016)]%
        {lin2016log}
\bibfield{author}{\bibinfo{person}{Qingwei Lin}, \bibinfo{person}{Hongyu
  Zhang}, \bibinfo{person}{Jian-Guang Lou}, \bibinfo{person}{Yu Zhang}, {and}
  \bibinfo{person}{Xuewei Chen}.} \bibinfo{year}{2016}\natexlab{}.
\newblock \showarticletitle{Log clustering based problem identification for
  online service systems}. In \bibinfo{booktitle}{\emph{Proceedings of the 38th
  International Conference on Software Engineering Companion}}.
  \bibinfo{pages}{102--111}.
\newblock


\bibitem[Lin et~al\mbox{.}(2018b)]%
        {lin2018facgraph}
\bibfield{author}{\bibinfo{person}{Weilan Lin}, \bibinfo{person}{Meng Ma},
  \bibinfo{person}{Disheng Pan}, {and} \bibinfo{person}{Ping Wang}.}
  \bibinfo{year}{2018}\natexlab{b}.
\newblock \showarticletitle{Facgraph: Frequent anomaly correlation graph mining
  for root cause diagnose in micro-service architecture}. In
  \bibinfo{booktitle}{\emph{2018 IEEE 37th International Performance Computing
  and Communications Conference (IPCCC)}}. IEEE, \bibinfo{pages}{1--8}.
\newblock


\bibitem[Liu et~al\mbox{.}(2021)]%
        {liu2021microhecl}
\bibfield{author}{\bibinfo{person}{Dewei Liu}, \bibinfo{person}{Chuan He},
  \bibinfo{person}{Xin Peng}, \bibinfo{person}{Fan Lin},
  \bibinfo{person}{Chenxi Zhang}, \bibinfo{person}{Shengfang Gong},
  \bibinfo{person}{Ziang Li}, \bibinfo{person}{Jiayu Ou}, {and}
  \bibinfo{person}{Zheshun Wu}.} \bibinfo{year}{2021}\natexlab{}.
\newblock \showarticletitle{Microhecl: High-efficient root cause localization
  in large-scale microservice systems}. In \bibinfo{booktitle}{\emph{2021
  IEEE/ACM 43rd International Conference on Software Engineering: Software
  Engineering in Practice (ICSE-SEIP)}}. IEEE, \bibinfo{pages}{338--347}.
\newblock


\bibitem[Liu et~al\mbox{.}(2022)]%
        {liu2022microcbr}
\bibfield{author}{\bibinfo{person}{Fengrui Liu}, \bibinfo{person}{Yang Wang},
  \bibinfo{person}{Zhenyu Li}, \bibinfo{person}{Rui Ren},
  \bibinfo{person}{Hongtao Guan}, \bibinfo{person}{Xian Yu},
  \bibinfo{person}{Xiaofan Chen}, {and} \bibinfo{person}{Gaogang Xie}.}
  \bibinfo{year}{2022}\natexlab{}.
\newblock \showarticletitle{MicroCBR: Case-Based Reasoning on Spatio-temporal
  Fault Knowledge Graph for Microservices Troubleshooting}. In
  \bibinfo{booktitle}{\emph{International Conference on Case-Based Reasoning}}.
  Springer, \bibinfo{pages}{224--239}.
\newblock


\bibitem[Liu et~al\mbox{.}(2020)]%
        {liu2020unsupervised}
\bibfield{author}{\bibinfo{person}{Ping Liu}, \bibinfo{person}{Haowen Xu},
  \bibinfo{person}{Qianyu Ouyang}, \bibinfo{person}{Rui Jiao},
  \bibinfo{person}{Zhekang Chen}, \bibinfo{person}{Shenglin Zhang},
  \bibinfo{person}{Jiahai Yang}, \bibinfo{person}{Linlin Mo},
  \bibinfo{person}{Jice Zeng}, \bibinfo{person}{Wenman Xue}, {et~al\mbox{.}}}
  \bibinfo{year}{2020}\natexlab{}.
\newblock \showarticletitle{Unsupervised detection of microservice trace
  anomalies through service-level deep bayesian networks}. In
  \bibinfo{booktitle}{\emph{2020 IEEE 31st International Symposium on Software
  Reliability Engineering (ISSRE)}}. IEEE, \bibinfo{pages}{48--58}.
\newblock


\bibitem[{Lseino}(2024)]%
        {lseino_onlineboutique}
\bibfield{author}{\bibinfo{person}{{Lseino}}.} \bibinfo{year}{2024}\natexlab{}.
\newblock \bibinfo{title}{lseino/onlineboutique}.
\newblock
\newblock
\urldef\tempurl%
\url{https://github.com/lseino/onlineboutique}
\showURL{%
\tempurl}
\newblock
\shownote{Accessed: 2024-05-26}.


\bibitem[Lu et~al\mbox{.}(2017)]%
        {lu2017log}
\bibfield{author}{\bibinfo{person}{Siyang Lu}, \bibinfo{person}{BingBing Rao},
  \bibinfo{person}{Xiang Wei}, \bibinfo{person}{Byungchul Tak},
  \bibinfo{person}{Long Wang}, {and} \bibinfo{person}{Liqiang Wang}.}
  \bibinfo{year}{2017}\natexlab{}.
\newblock \showarticletitle{Log-based abnormal task detection and root cause
  analysis for spark}. In \bibinfo{booktitle}{\emph{2017 IEEE International
  Conference on Web Services (ICWS)}}. IEEE, \bibinfo{pages}{389--396}.
\newblock


\bibitem[Ma et~al\mbox{.}(2019)]%
        {ma2019ms}
\bibfield{author}{\bibinfo{person}{Meng Ma}, \bibinfo{person}{Weilan Lin},
  \bibinfo{person}{Disheng Pan}, {and} \bibinfo{person}{Ping Wang}.}
  \bibinfo{year}{2019}\natexlab{}.
\newblock \showarticletitle{Ms-rank: Multi-metric and self-adaptive root cause
  diagnosis for microservice applications}. In \bibinfo{booktitle}{\emph{2019
  IEEE International Conference on Web Services (ICWS)}}. IEEE,
  \bibinfo{pages}{60--67}.
\newblock


\bibitem[Ma et~al\mbox{.}(2020a)]%
        {ma2020automap}
\bibfield{author}{\bibinfo{person}{Meng Ma}, \bibinfo{person}{Jingmin Xu},
  \bibinfo{person}{Yuan Wang}, \bibinfo{person}{Pengfei Chen},
  \bibinfo{person}{Zonghua Zhang}, {and} \bibinfo{person}{Ping Wang}.}
  \bibinfo{year}{2020}\natexlab{a}.
\newblock \showarticletitle{Automap: Diagnose your microservice-based web
  applications automatically}. In \bibinfo{booktitle}{\emph{Proceedings of The
  Web Conference 2020}}. \bibinfo{pages}{246--258}.
\newblock


\bibitem[Ma et~al\mbox{.}(2020b)]%
        {ma2020diagnosing}
\bibfield{author}{\bibinfo{person}{Minghua Ma}, \bibinfo{person}{Zheng Yin},
  \bibinfo{person}{Shenglin Zhang}, \bibinfo{person}{Sheng Wang},
  \bibinfo{person}{Christopher Zheng}, \bibinfo{person}{Xinhao Jiang},
  \bibinfo{person}{Hanwen Hu}, \bibinfo{person}{Cheng Luo},
  \bibinfo{person}{Yilin Li}, \bibinfo{person}{Nengjun Qiu}, {et~al\mbox{.}}}
  \bibinfo{year}{2020}\natexlab{b}.
\newblock \showarticletitle{Diagnosing root causes of intermittent slow queries
  in cloud databases}.
\newblock \bibinfo{journal}{\emph{Proceedings of the VLDB Endowment}}
  \bibinfo{volume}{13}, \bibinfo{number}{8} (\bibinfo{year}{2020}),
  \bibinfo{pages}{1176--1189}.
\newblock


\bibitem[Maeyens et~al\mbox{.}(2020)]%
        {maeyens2020process}
\bibfield{author}{\bibinfo{person}{Jonas Maeyens}, \bibinfo{person}{Annemie
  Vorstermans}, {and} \bibinfo{person}{Mathias Verbeke}.}
  \bibinfo{year}{2020}\natexlab{}.
\newblock \showarticletitle{Process mining on machine event logs for profiling
  abnormal behaviour and root cause analysis}.
\newblock \bibinfo{journal}{\emph{Annals of Telecommunications}}
  \bibinfo{volume}{75}, \bibinfo{number}{9} (\bibinfo{year}{2020}),
  \bibinfo{pages}{563--572}.
\newblock


\bibitem[Mandal et~al\mbox{.}(2021)]%
        {mandal2021localization}
\bibfield{author}{\bibinfo{person}{Atri Mandal}, \bibinfo{person}{Qing Wang},
  {and} \bibinfo{person}{Amit Paradkar}.} \bibinfo{year}{2021}\natexlab{}.
\newblock \showarticletitle{Localization of Operational Faults in Cloud
  Applications by Mining Causal Dependencies in Logs Using Golden Signals}. In
  \bibinfo{booktitle}{\emph{Service-Oriented Computing--ICSOC 2020 Workshops:
  AIOps, CFTIC, STRAPS, AI-PA, AI-IOTS, and Satellite Events, Dubai, United
  Arab Emirates, December 14--17, 2020, Proceedings}},
  Vol.~\bibinfo{volume}{12632}. Springer Nature, \bibinfo{pages}{137}.
\newblock


\bibitem[Mariani et~al\mbox{.}(2018)]%
        {mariani2018localizing}
\bibfield{author}{\bibinfo{person}{Leonardo Mariani}, \bibinfo{person}{Cristina
  Monni}, \bibinfo{person}{Mauro Pezz{\'e}}, \bibinfo{person}{Oliviero
  Riganelli}, {and} \bibinfo{person}{Rui Xin}.}
  \bibinfo{year}{2018}\natexlab{}.
\newblock \showarticletitle{Localizing faults in cloud systems}. In
  \bibinfo{booktitle}{\emph{2018 IEEE 11th International Conference on Software
  Testing, Verification and Validation (ICST)}}. IEEE,
  \bibinfo{pages}{262--273}.
\newblock


\bibitem[Marwede et~al\mbox{.}(2009)]%
        {marwede2009automatic}
\bibfield{author}{\bibinfo{person}{Nina Marwede}, \bibinfo{person}{Matthias
  Rohr}, \bibinfo{person}{Andr{\'e} van Hoorn}, {and} \bibinfo{person}{Wilhelm
  Hasselbring}.} \bibinfo{year}{2009}\natexlab{}.
\newblock \showarticletitle{Automatic failure diagnosis support in distributed
  large-scale software systems based on timing behavior anomaly correlation}.
  In \bibinfo{booktitle}{\emph{2009 13th European Conference on Software
  Maintenance and Reengineering}}. IEEE, \bibinfo{pages}{47--58}.
\newblock


\bibitem[{Mashable}(2023)]%
        {Mashable2023}
\bibfield{author}{\bibinfo{person}{{Mashable}}.}
  \bibinfo{year}{2023}\natexlab{}.
\newblock \bibinfo{booktitle}{\emph{{ChatGPT was down. What we know about the
  major outage.}}}
\newblock


\bibitem[Meng et~al\mbox{.}(2020)]%
        {meng2020localizing}
\bibfield{author}{\bibinfo{person}{Yuan Meng}, \bibinfo{person}{Shenglin
  Zhang}, \bibinfo{person}{Yongqian Sun}, \bibinfo{person}{Ruru Zhang},
  \bibinfo{person}{Zhilong Hu}, \bibinfo{person}{Yiyin Zhang},
  \bibinfo{person}{Chenyang Jia}, \bibinfo{person}{Zhaogang Wang}, {and}
  \bibinfo{person}{Dan Pei}.} \bibinfo{year}{2020}\natexlab{}.
\newblock \showarticletitle{Localizing failure root causes in a microservice
  through causality inference}. In \bibinfo{booktitle}{\emph{2020 IEEE/ACM 28th
  International Symposium on Quality of Service (IWQoS)}}. IEEE,
  \bibinfo{pages}{1--10}.
\newblock


\bibitem[Mi et~al\mbox{.}(2012)]%
        {mi2012localizing}
\bibfield{author}{\bibinfo{person}{HaiBo Mi}, \bibinfo{person}{HuaiMin Wang},
  \bibinfo{person}{YangFan Zhou}, \bibinfo{person}{Michael~R Lyu}, {and}
  \bibinfo{person}{Hua Cai}.} \bibinfo{year}{2012}\natexlab{}.
\newblock \showarticletitle{Localizing root causes of performance anomalies in
  cloud computing systems by analyzing request trace logs}.
\newblock \bibinfo{journal}{\emph{Science China Information Sciences}}
  \bibinfo{volume}{55} (\bibinfo{year}{2012}), \bibinfo{pages}{2757--2773}.
\newblock


\bibitem[{Ministry of Industry and Information Technology of China
  (Miit)}(2023)]%
        {Miit2023notice}
\bibfield{author}{\bibinfo{person}{{Ministry of Industry and Information
  Technology of China (Miit)}}.} \bibinfo{year}{2023}\natexlab{}.
\newblock \bibinfo{booktitle}{\emph{{Tencent Guangzhou Availability Zone Fault
  Incident}}}.
\newblock


\bibitem[Nedelkoski et~al\mbox{.}(2019)]%
        {nedelkoski2019anomaly}
\bibfield{author}{\bibinfo{person}{Sasho Nedelkoski}, \bibinfo{person}{Jorge
  Cardoso}, {and} \bibinfo{person}{Odej Kao}.} \bibinfo{year}{2019}\natexlab{}.
\newblock \showarticletitle{Anomaly detection from system tracing data using
  multimodal deep learning}. In \bibinfo{booktitle}{\emph{2019 IEEE 12th
  International Conference on Cloud Computing (CLOUD)}}. IEEE,
  \bibinfo{pages}{179--186}.
\newblock


\bibitem[Nguyen et~al\mbox{.}(2013)]%
        {nguyen2013fchain}
\bibfield{author}{\bibinfo{person}{Hiep Nguyen}, \bibinfo{person}{Zhiming
  Shen}, \bibinfo{person}{Yongmin Tan}, {and} \bibinfo{person}{Xiaohui Gu}.}
  \bibinfo{year}{2013}\natexlab{}.
\newblock \showarticletitle{Fchain: Toward black-box online fault localization
  for cloud systems}. In \bibinfo{booktitle}{\emph{2013 IEEE 33rd International
  Conference on Distributed Computing Systems}}. IEEE, \bibinfo{pages}{21--30}.
\newblock


\bibitem[Nguyen et~al\mbox{.}(2011)]%
        {nguyen2011pal}
\bibfield{author}{\bibinfo{person}{Hiep Nguyen}, \bibinfo{person}{Yongmin Tan},
  {and} \bibinfo{person}{Xiaohui Gu}.} \bibinfo{year}{2011}\natexlab{}.
\newblock \showarticletitle{Pal: Propagation-aware a nomaly l ocalization for
  cloud hosted distributed applications}.
\newblock In \bibinfo{booktitle}{\emph{Managing Large-scale Systems via the
  Analysis of System Logs and the Application of Machine Learning Techniques}}.
  \bibinfo{pages}{1--8}.
\newblock


\bibitem[Nie et~al\mbox{.}(2016)]%
        {nie2016mining}
\bibfield{author}{\bibinfo{person}{Xiaohui Nie}, \bibinfo{person}{Youjian
  Zhao}, \bibinfo{person}{Kaixin Sui}, \bibinfo{person}{Dan Pei},
  \bibinfo{person}{Yu Chen}, {and} \bibinfo{person}{Xianping Qu}.}
  \bibinfo{year}{2016}\natexlab{}.
\newblock \showarticletitle{Mining causality graph for automatic web-based
  service diagnosis}. In \bibinfo{booktitle}{\emph{2016 IEEE 35th International
  Performance Computing and Communications Conference (IPCCC)}}. IEEE,
  \bibinfo{pages}{1--8}.
\newblock


\bibitem[Notaro et~al\mbox{.}(2023)]%
        {notaro2023logrule}
\bibfield{author}{\bibinfo{person}{Paolo Notaro}, \bibinfo{person}{Soroush
  Haeri}, \bibinfo{person}{Jorge Cardoso}, {and} \bibinfo{person}{Michael
  Gerndt}.} \bibinfo{year}{2023}\natexlab{}.
\newblock \showarticletitle{LogRule: Efficient structured log mining for root
  cause analysis}.
\newblock \bibinfo{journal}{\emph{IEEE Transactions on Network and Service
  Management}} (\bibinfo{year}{2023}).
\newblock


\bibitem[{OpenAI}(2024)]%
        {OpenAI2024}
\bibfield{author}{\bibinfo{person}{{OpenAI}}.} \bibinfo{year}{2024}\natexlab{}.
\newblock \bibinfo{booktitle}{\emph{{Incidents of OpenAI}}}.
\newblock


\bibitem[{opensource-socialnetwork}(2024)]%
        {opensource_socialnetwork}
\bibfield{author}{\bibinfo{person}{{opensource-socialnetwork}}.}
  \bibinfo{year}{2024}\natexlab{}.
\newblock \bibinfo{title}{opensource-socialnetwork/opensource-socialnetwork}.
\newblock
\newblock
\urldef\tempurl%
\url{https://github.com/opensource-socialnetwork/opensource-socialnetwork}
\showURL{%
\tempurl}
\newblock
\shownote{Accessed: 2024-05-26}.


\bibitem[Qiu et~al\mbox{.}(2020)]%
        {qiu2020causality}
\bibfield{author}{\bibinfo{person}{Juan Qiu}, \bibinfo{person}{Qingfeng Du},
  \bibinfo{person}{Kanglin Yin}, \bibinfo{person}{Shuang-Li Zhang}, {and}
  \bibinfo{person}{Chongshu Qian}.} \bibinfo{year}{2020}\natexlab{}.
\newblock \showarticletitle{A causality mining and knowledge graph based method
  of root cause diagnosis for performance anomaly in cloud applications}.
\newblock \bibinfo{journal}{\emph{Applied Sciences}} \bibinfo{volume}{10},
  \bibinfo{number}{6} (\bibinfo{year}{2020}), \bibinfo{pages}{2166}.
\newblock


\bibitem[Ren et~al\mbox{.}(2019)]%
        {ren2019root}
\bibfield{author}{\bibinfo{person}{Zhilei Ren}, \bibinfo{person}{Changlin Liu},
  \bibinfo{person}{Xusheng Xiao}, \bibinfo{person}{He Jiang}, {and}
  \bibinfo{person}{Tao Xie}.} \bibinfo{year}{2019}\natexlab{}.
\newblock \showarticletitle{Root cause localization for unreproducible builds
  via causality analysis over system call tracing}. In
  \bibinfo{booktitle}{\emph{2019 34th IEEE/ACM International Conference on
  Automated Software Engineering (ASE)}}. IEEE, \bibinfo{pages}{527--538}.
\newblock


\bibitem[Rosenberg and Moonen(2020)]%
        {rosenberg2020spectrum}
\bibfield{author}{\bibinfo{person}{Carl~Martin Rosenberg} {and}
  \bibinfo{person}{Leon Moonen}.} \bibinfo{year}{2020}\natexlab{}.
\newblock \showarticletitle{Spectrum-based log diagnosis}. In
  \bibinfo{booktitle}{\emph{Proceedings of the 14th ACM/IEEE International
  Symposium on Empirical Software Engineering and Measurement (ESEM)}}.
  \bibinfo{pages}{1--12}.
\newblock


\bibitem[Roy et~al\mbox{.}(2024)]%
        {roy2024exploring}
\bibfield{author}{\bibinfo{person}{Devjeet Roy}, \bibinfo{person}{Xuchao
  Zhang}, \bibinfo{person}{Rashi Bhave}, \bibinfo{person}{Chetan Bansal},
  \bibinfo{person}{Pedro Las-Casas}, \bibinfo{person}{Rodrigo Fonseca}, {and}
  \bibinfo{person}{Saravan Rajmohan}.} \bibinfo{year}{2024}\natexlab{}.
\newblock \showarticletitle{Exploring LLM-based Agents for Root Cause
  Analysis}.
\newblock \bibinfo{journal}{\emph{arXiv preprint arXiv:2403.04123}}
  (\bibinfo{year}{2024}).
\newblock


\bibitem[Saha and Hoi(2022)]%
        {saha2022mining}
\bibfield{author}{\bibinfo{person}{Amrita Saha} {and}
  \bibinfo{person}{Steven~CH Hoi}.} \bibinfo{year}{2022}\natexlab{}.
\newblock \showarticletitle{Mining root cause knowledge from cloud service
  incident investigations for AIOps}. In \bibinfo{booktitle}{\emph{Proceedings
  of the 44th International Conference on Software Engineering: Software
  Engineering in Practice}}. \bibinfo{pages}{197--206}.
\newblock


\bibitem[Samir and Pahl(2019)]%
        {samir2019dla}
\bibfield{author}{\bibinfo{person}{Areeg Samir} {and} \bibinfo{person}{Claus
  Pahl}.} \bibinfo{year}{2019}\natexlab{}.
\newblock \showarticletitle{Dla: Detecting and localizing anomalies in
  containerized microservice architectures using markov models}. In
  \bibinfo{booktitle}{\emph{2019 7th International Conference on Future
  Internet of Things and Cloud (FiCloud)}}. IEEE, \bibinfo{pages}{205--213}.
\newblock


\bibitem[Sarda(2023)]%
        {sarda2023leveraging}
\bibfield{author}{\bibinfo{person}{Komal Sarda}.}
  \bibinfo{year}{2023}\natexlab{}.
\newblock \showarticletitle{Leveraging Large Language Models for
  Auto-remediation in Microservices Architecture}. In
  \bibinfo{booktitle}{\emph{2023 IEEE International Conference on Autonomic
  Computing and Self-Organizing Systems Companion (ACSOS-C)}}. IEEE,
  \bibinfo{pages}{16--18}.
\newblock


\bibitem[Sarda et~al\mbox{.}(2023)]%
        {sarda2023adarma}
\bibfield{author}{\bibinfo{person}{Komal Sarda}, \bibinfo{person}{Zakeya
  Namrud}, \bibinfo{person}{Raphael Rouf}, \bibinfo{person}{Harit Ahuja},
  \bibinfo{person}{Mohammadreza Rasolroveicy}, \bibinfo{person}{Marin Litoiu},
  \bibinfo{person}{Larisa Shwartz}, {and} \bibinfo{person}{Ian Watts}.}
  \bibinfo{year}{2023}\natexlab{}.
\newblock \showarticletitle{Adarma auto-detection and auto-remediation of
  microservice anomalies by leveraging large language models}. In
  \bibinfo{booktitle}{\emph{Proceedings of the 33rd Annual International
  Conference on Computer Science and Software Engineering}}.
  \bibinfo{pages}{200--205}.
\newblock


\bibitem[Setiawan et~al\mbox{.}(2020)]%
        {setiawan2020gwad}
\bibfield{author}{\bibinfo{person}{Wiliam Setiawan}, \bibinfo{person}{Yohen
  Thounaojam}, {and} \bibinfo{person}{Apurva Narayan}.}
  \bibinfo{year}{2020}\natexlab{}.
\newblock \showarticletitle{Gwad: Greedy workflow graph anomaly detection
  framework for system traces}. In \bibinfo{booktitle}{\emph{2020 IEEE
  International Conference on systems, man, and Cybernetics (SMC)}}. IEEE,
  \bibinfo{pages}{2790--2796}.
\newblock


\bibitem[Shan et~al\mbox{.}(2019)]%
        {shan2019diagnosis}
\bibfield{author}{\bibinfo{person}{Huasong Shan}, \bibinfo{person}{Yuan Chen},
  \bibinfo{person}{Haifeng Liu}, \bibinfo{person}{Yunpeng Zhang},
  \bibinfo{person}{Xiao Xiao}, \bibinfo{person}{Xiaofeng He},
  \bibinfo{person}{Min Li}, {and} \bibinfo{person}{Wei Ding}.}
  \bibinfo{year}{2019}\natexlab{}.
\newblock \showarticletitle{\(\varepsilon\)-diagnosis: Unsupervised and
  real-time diagnosis of small-window long-tail latency in large-scale
  microservice platforms}. In \bibinfo{booktitle}{\emph{The World Wide Web
  Conference}}. \bibinfo{pages}{3215--3222}.
\newblock


\bibitem[Shetty et~al\mbox{.}(2022a)]%
        {shetty2022softner}
\bibfield{author}{\bibinfo{person}{Manish Shetty}, \bibinfo{person}{Chetan
  Bansal}, \bibinfo{person}{Sumit Kumar}, \bibinfo{person}{Nikitha Rao}, {and}
  \bibinfo{person}{Nachiappan Nagappan}.} \bibinfo{year}{2022}\natexlab{a}.
\newblock \showarticletitle{SoftNER: Mining knowledge graphs from cloud
  incidents}.
\newblock \bibinfo{journal}{\emph{Empirical Software Engineering}}
  \bibinfo{volume}{27}, \bibinfo{number}{4} (\bibinfo{year}{2022}),
  \bibinfo{pages}{93}.
\newblock


\bibitem[Shetty et~al\mbox{.}(2022b)]%
        {shetty2022autotsg}
\bibfield{author}{\bibinfo{person}{Manish Shetty}, \bibinfo{person}{Chetan
  Bansal}, \bibinfo{person}{Sai~Pramod Upadhyayula}, \bibinfo{person}{Arjun
  Radhakrishna}, {and} \bibinfo{person}{Anurag Gupta}.}
  \bibinfo{year}{2022}\natexlab{b}.
\newblock \showarticletitle{AutoTSG: learning and synthesis for incident
  troubleshooting}. In \bibinfo{booktitle}{\emph{Proceedings of the 30th ACM
  Joint European Software Engineering Conference and Symposium on the
  Foundations of Software Engineering}}. \bibinfo{pages}{1477--1488}.
\newblock


\bibitem[Siebert(2023)]%
        {siebert2023applications}
\bibfield{author}{\bibinfo{person}{Julien Siebert}.}
  \bibinfo{year}{2023}\natexlab{}.
\newblock \showarticletitle{Applications of statistical causal inference in
  software engineering}.
\newblock \bibinfo{journal}{\emph{Information and Software Technology}}
  (\bibinfo{year}{2023}), \bibinfo{pages}{107198}.
\newblock


\bibitem[Soldani and Brogi(2022)]%
        {soldani2022anomaly}
\bibfield{author}{\bibinfo{person}{Jacopo Soldani} {and}
  \bibinfo{person}{Antonio Brogi}.} \bibinfo{year}{2022}\natexlab{}.
\newblock \showarticletitle{Anomaly detection and failure root cause analysis
  in (micro) service-based cloud applications: A survey}.
\newblock \bibinfo{journal}{\emph{ACM Computing Surveys (CSUR)}}
  \bibinfo{volume}{55}, \bibinfo{number}{3} (\bibinfo{year}{2022}),
  \bibinfo{pages}{1--39}.
\newblock


\bibitem[Soldani et~al\mbox{.}(2022)]%
        {soldani2022failure}
\bibfield{author}{\bibinfo{person}{Jacopo Soldani}, \bibinfo{person}{Stefano
  Forti}, {and} \bibinfo{person}{Antonio Brogi}.}
  \bibinfo{year}{2022}\natexlab{}.
\newblock \showarticletitle{Failure root cause analysis for microservices,
  explained}. In \bibinfo{booktitle}{\emph{IFIP International Conference on
  Distributed Applications and Interoperable Systems}}. Springer,
  \bibinfo{pages}{74--91}.
\newblock


\bibitem[Sol{\'e} et~al\mbox{.}(2017)]%
        {sole2017survey}
\bibfield{author}{\bibinfo{person}{Marc Sol{\'e}}, \bibinfo{person}{Victor
  Munt{\'e}s-Mulero}, \bibinfo{person}{Annie~Ibrahim Rana}, {and}
  \bibinfo{person}{Giovani Estrada}.} \bibinfo{year}{2017}\natexlab{}.
\newblock \showarticletitle{Survey on models and techniques for root-cause
  analysis}.
\newblock \bibinfo{journal}{\emph{arXiv preprint arXiv:1701.08546}}
  (\bibinfo{year}{2017}).
\newblock


\bibitem[Soualhia and Wuhib(2022)]%
        {soualhia2022automated}
\bibfield{author}{\bibinfo{person}{Mbarka Soualhia} {and}
  \bibinfo{person}{Fetahi Wuhib}.} \bibinfo{year}{2022}\natexlab{}.
\newblock \showarticletitle{Automated traces-based anomaly detection and root
  cause analysis in cloud platforms}. In \bibinfo{booktitle}{\emph{2022 IEEE
  International Conference on Cloud Engineering (IC2E)}}. IEEE,
  \bibinfo{pages}{253--260}.
\newblock


\bibitem[Sun et~al\mbox{.}(2021)]%
        {sun2021fault}
\bibfield{author}{\bibinfo{person}{Yindong Sun}, \bibinfo{person}{Longjun
  Zhao}, \bibinfo{person}{Zhen Wang}, \bibinfo{person}{Dandan Cui},
  \bibinfo{person}{Yang Yang}, {and} \bibinfo{person}{Zhipeng Gao}.}
  \bibinfo{year}{2021}\natexlab{}.
\newblock \showarticletitle{Fault Root Rank Algorithm Based on Random Walk
  Mechanism in Fault Knowledge Graph}. In \bibinfo{booktitle}{\emph{2021 IEEE
  International Symposium on Broadband Multimedia Systems and Broadcasting
  (BMSB)}}. IEEE, \bibinfo{pages}{1--6}.
\newblock


\bibitem[Suriadi et~al\mbox{.}(2013)]%
        {suriadi2013root}
\bibfield{author}{\bibinfo{person}{Suriadi Suriadi}, \bibinfo{person}{Chun
  Ouyang}, \bibinfo{person}{Wil~MP van~der Aalst}, {and}
  \bibinfo{person}{Arthur~HM ter Hofstede}.} \bibinfo{year}{2013}\natexlab{}.
\newblock \showarticletitle{Root cause analysis with enriched process logs}. In
  \bibinfo{booktitle}{\emph{Business Process Management Workshops: BPM 2012
  International Workshops, Tallinn, Estonia, September 3, 2012. Revised Papers
  10}}. Springer, \bibinfo{pages}{174--186}.
\newblock


\bibitem[Tak et~al\mbox{.}(2016)]%
        {tak2016logan}
\bibfield{author}{\bibinfo{person}{Byung~Chul Tak}, \bibinfo{person}{Shu Tao},
  \bibinfo{person}{Lin Yang}, \bibinfo{person}{Chao Zhu}, {and}
  \bibinfo{person}{Yaoping Ruan}.} \bibinfo{year}{2016}\natexlab{}.
\newblock \showarticletitle{Logan: Problem diagnosis in the cloud using
  log-based reference models}. In \bibinfo{booktitle}{\emph{2016 IEEE
  International Conference on Cloud Engineering (IC2E)}}. IEEE,
  \bibinfo{pages}{62--67}.
\newblock


\bibitem[Tan et~al\mbox{.}(2017)]%
        {tan2017hwlog}
\bibfield{author}{\bibinfo{person}{Tun-Zi Tan}, \bibinfo{person}{Xing-Chen
  Zhang}, \bibinfo{person}{Sui-Xiang Gao}, \bibinfo{person}{Wen-Guo Yang},
  \bibinfo{person}{Yue-Zhong Song}, {and} \bibinfo{person}{Cheng-Yong Lin}.}
  \bibinfo{year}{2017}\natexlab{}.
\newblock \showarticletitle{HWLog Analysis: A Tool for Routers’ Syslog
  Anomaly Detection and Root Causes Diagnosis}. In
  \bibinfo{booktitle}{\emph{Artificial Intelligence Science and Technology:
  Proceedings of the 2016 International Conference (AIST2016)}}. World
  Scientific, \bibinfo{pages}{799--806}.
\newblock


\bibitem[Thalheim et~al\mbox{.}(2017)]%
        {thalheim2017sieve}
\bibfield{author}{\bibinfo{person}{J{\"o}rg Thalheim}, \bibinfo{person}{Antonio
  Rodrigues}, \bibinfo{person}{Istemi~Ekin Akkus}, \bibinfo{person}{Pramod
  Bhatotia}, \bibinfo{person}{Ruichuan Chen}, \bibinfo{person}{Bimal
  Viswanath}, \bibinfo{person}{Lei Jiao}, {and} \bibinfo{person}{Christof
  Fetzer}.} \bibinfo{year}{2017}\natexlab{}.
\newblock \showarticletitle{Sieve: Actionable insights from monitored metrics
  in distributed systems}. In \bibinfo{booktitle}{\emph{Proceedings of the 18th
  ACM/IFIP/USENIX Middleware Conference}}. \bibinfo{pages}{14--27}.
\newblock


\bibitem[{Theregister}(2023)]%
        {GoogleDrive2023}
\bibfield{author}{\bibinfo{person}{{Theregister}}.}
  \bibinfo{year}{2023}\natexlab{}.
\newblock \bibinfo{booktitle}{\emph{{Google Drive misplaces months' worth of
  customer files}}}.
\newblock


\bibitem[Thurrott(2024)]%
        {Microsoft2024}
\bibfield{author}{\bibinfo{person}{Thurrott}.} \bibinfo{year}{2024}\natexlab{}.
\newblock \bibinfo{booktitle}{\emph{{Microsoft Bing Outage is Impacting
  Copilot, DuckDuckGo, And Other Services}}}.
\newblock


\bibitem[UniSuper(2024)]%
        {GoogleCloud2024}
\bibfield{author}{\bibinfo{person}{UniSuper}.} \bibinfo{year}{2024}\natexlab{}.
\newblock \bibinfo{booktitle}{\emph{{A joint statement from UniSuper CEO Peter
  Chun, and Google Cloud CEO, Thomas Kurian}}}.
\newblock


\bibitem[Wang et~al\mbox{.}(2023a)]%
        {wang2023incremental}
\bibfield{author}{\bibinfo{person}{Dongjie Wang}, \bibinfo{person}{Zhengzhang
  Chen}, \bibinfo{person}{Yanjie Fu}, \bibinfo{person}{Yanchi Liu}, {and}
  \bibinfo{person}{Haifeng Chen}.} \bibinfo{year}{2023}\natexlab{a}.
\newblock \showarticletitle{Incremental causal graph learning for online root
  cause analysis}. In \bibinfo{booktitle}{\emph{Proceedings of the 29th ACM
  SIGKDD Conference on Knowledge Discovery and Data Mining}}.
  \bibinfo{pages}{2269--2278}.
\newblock


\bibitem[Wang et~al\mbox{.}(2023b)]%
        {wang2023hierarchical}
\bibfield{author}{\bibinfo{person}{Dongjie Wang}, \bibinfo{person}{Zhengzhang
  Chen}, \bibinfo{person}{Jingchao Ni}, \bibinfo{person}{Liang Tong},
  \bibinfo{person}{Zheng Wang}, \bibinfo{person}{Yanjie Fu}, {and}
  \bibinfo{person}{Haifeng Chen}.} \bibinfo{year}{2023}\natexlab{b}.
\newblock \showarticletitle{Hierarchical graph neural networks for causal
  discovery and root cause localization}.
\newblock \bibinfo{journal}{\emph{arXiv preprint arXiv:2302.01987}}
  (\bibinfo{year}{2023}).
\newblock


\bibitem[Wang et~al\mbox{.}(2021)]%
        {wang2021groot}
\bibfield{author}{\bibinfo{person}{Hanzhang Wang}, \bibinfo{person}{Zhengkai
  Wu}, \bibinfo{person}{Huai Jiang}, \bibinfo{person}{Yichao Huang},
  \bibinfo{person}{Jiamu Wang}, \bibinfo{person}{Selcuk Kopru}, {and}
  \bibinfo{person}{Tao Xie}.} \bibinfo{year}{2021}\natexlab{}.
\newblock \showarticletitle{Groot: An event-graph-based approach for root cause
  analysis in industrial settings}. In \bibinfo{booktitle}{\emph{2021 36th
  IEEE/ACM International Conference on Automated Software Engineering (ASE)}}.
  IEEE, \bibinfo{pages}{419--429}.
\newblock


\bibitem[Wang et~al\mbox{.}(2020)]%
        {wang2020root}
\bibfield{author}{\bibinfo{person}{Lingzhi Wang}, \bibinfo{person}{Nengwen
  Zhao}, \bibinfo{person}{Junjie Chen}, \bibinfo{person}{Pinnong Li},
  \bibinfo{person}{Wenchi Zhang}, {and} \bibinfo{person}{Kaixin Sui}.}
  \bibinfo{year}{2020}\natexlab{}.
\newblock \showarticletitle{Root-cause metric location for microservice systems
  via log anomaly detection}. In \bibinfo{booktitle}{\emph{2020 IEEE
  international conference on web services (ICWS)}}. IEEE,
  \bibinfo{pages}{142--150}.
\newblock


\bibitem[Wang et~al\mbox{.}(2018)]%
        {wang2018cloudranger}
\bibfield{author}{\bibinfo{person}{Ping Wang}, \bibinfo{person}{Jingmin Xu},
  \bibinfo{person}{Meng Ma}, \bibinfo{person}{Weilan Lin},
  \bibinfo{person}{Disheng Pan}, \bibinfo{person}{Yuan Wang}, {and}
  \bibinfo{person}{Pengfei Chen}.} \bibinfo{year}{2018}\natexlab{}.
\newblock \showarticletitle{Cloudranger: Root cause identification for cloud
  native systems}. In \bibinfo{booktitle}{\emph{2018 18th IEEE/ACM
  International Symposium on Cluster, Cloud and Grid Computing (CCGRID)}}.
  IEEE, \bibinfo{pages}{492--502}.
\newblock


\bibitem[Wang et~al\mbox{.}(2023c)]%
        {wang2023rcagent}
\bibfield{author}{\bibinfo{person}{Zefan Wang}, \bibinfo{person}{Zichuan Liu},
  \bibinfo{person}{Yingying Zhang}, \bibinfo{person}{Aoxiao Zhong},
  \bibinfo{person}{Lunting Fan}, \bibinfo{person}{Lingfei Wu}, {and}
  \bibinfo{person}{Qingsong Wen}.} \bibinfo{year}{2023}\natexlab{c}.
\newblock \showarticletitle{RCAgent: Cloud Root Cause Analysis by Autonomous
  Agents with Tool-Augmented Large Language Models}.
\newblock \bibinfo{journal}{\emph{arXiv preprint arXiv:2310.16340}}
  (\bibinfo{year}{2023}).
\newblock


\bibitem[Wanga(2018)]%
        {wanga2018ladra}
\bibfield{author}{\bibinfo{person}{Liqiang Wanga}.}
  \bibinfo{year}{2018}\natexlab{}.
\newblock \showarticletitle{LADRA: Log-Based Abnormal Task Detection and
  Root-Cause Analysis in Big Data Processing with Spark}.
\newblock  (\bibinfo{year}{2018}).
\newblock


\bibitem[Weng et~al\mbox{.}(2018)]%
        {weng2018root}
\bibfield{author}{\bibinfo{person}{Jianping Weng}, \bibinfo{person}{Jessie~Hui
  Wang}, \bibinfo{person}{Jiahai Yang}, {and} \bibinfo{person}{Yang Yang}.}
  \bibinfo{year}{2018}\natexlab{}.
\newblock \showarticletitle{Root cause analysis of anomalies of multitier
  services in public clouds}.
\newblock \bibinfo{journal}{\emph{IEEE/ACM Transactions on Networking}}
  \bibinfo{volume}{26}, \bibinfo{number}{4} (\bibinfo{year}{2018}),
  \bibinfo{pages}{1646--1659}.
\newblock


\bibitem[White et~al\mbox{.}(2021)]%
        {white2021mmrca}
\bibfield{author}{\bibinfo{person}{Gary White}, \bibinfo{person}{Jaroslaw
  Diuwe}, \bibinfo{person}{Erika Fonseca}, {and} \bibinfo{person}{Owen
  O’Brien}.} \bibinfo{year}{2021}\natexlab{}.
\newblock \showarticletitle{MMRCA: multimodal root cause analysis}. In
  \bibinfo{booktitle}{\emph{International Conference on Service-Oriented
  Computing}}. Springer, \bibinfo{pages}{177--189}.
\newblock


\bibitem[Wu et~al\mbox{.}(2020a)]%
        {wu2020performance}
\bibfield{author}{\bibinfo{person}{Li Wu}, \bibinfo{person}{Jasmin
  Bogatinovski}, \bibinfo{person}{Sasho Nedelkoski}, \bibinfo{person}{Johan
  Tordsson}, {and} \bibinfo{person}{Odej Kao}.}
  \bibinfo{year}{2020}\natexlab{a}.
\newblock \showarticletitle{Performance diagnosis in cloud microservices using
  deep learning}. In \bibinfo{booktitle}{\emph{International Conference on
  Service-Oriented Computing}}. Springer, \bibinfo{pages}{85--96}.
\newblock


\bibitem[Wu et~al\mbox{.}(2020b)]%
        {wu2020microrca}
\bibfield{author}{\bibinfo{person}{Li Wu}, \bibinfo{person}{Johan Tordsson},
  \bibinfo{person}{Erik Elmroth}, {and} \bibinfo{person}{Odej Kao}.}
  \bibinfo{year}{2020}\natexlab{b}.
\newblock \showarticletitle{Microrca: Root cause localization of performance
  issues in microservices}. In \bibinfo{booktitle}{\emph{NOMS 2020-2020
  IEEE/IFIP Network Operations and Management Symposium}}. IEEE,
  \bibinfo{pages}{1--9}.
\newblock


\bibitem[Xia et~al\mbox{.}(2022)]%
        {xia2022toward}
\bibfield{author}{\bibinfo{person}{Liqiao Xia}, \bibinfo{person}{Pai Zheng},
  \bibinfo{person}{Xinyu Li}, \bibinfo{person}{Robert~X Gao}, {and}
  \bibinfo{person}{Lihui Wang}.} \bibinfo{year}{2022}\natexlab{}.
\newblock \showarticletitle{Toward cognitive predictive maintenance: A survey
  of graph-based approaches}.
\newblock \bibinfo{journal}{\emph{Journal of Manufacturing Systems}}
  \bibinfo{volume}{64} (\bibinfo{year}{2022}), \bibinfo{pages}{107--120}.
\newblock


\bibitem[Xie et~al\mbox{.}(2023)]%
        {xie2023unsupervised}
\bibfield{author}{\bibinfo{person}{Zhe Xie}, \bibinfo{person}{Haowen Xu},
  \bibinfo{person}{Wenxiao Chen}, \bibinfo{person}{Wanxue Li},
  \bibinfo{person}{Huai Jiang}, \bibinfo{person}{Liangfei Su},
  \bibinfo{person}{Hanzhang Wang}, {and} \bibinfo{person}{Dan Pei}.}
  \bibinfo{year}{2023}\natexlab{}.
\newblock \showarticletitle{Unsupervised Anomaly Detection on Microservice
  Traces through Graph VAE}. In \bibinfo{booktitle}{\emph{Proceedings of the
  ACM Web Conference 2023}}. \bibinfo{pages}{2874--2884}.
\newblock


\bibitem[Xu et~al\mbox{.}(2017)]%
        {xu2017logdc}
\bibfield{author}{\bibinfo{person}{Jingmin Xu}, \bibinfo{person}{Pengfei Chen},
  \bibinfo{person}{Lin Yang}, \bibinfo{person}{Fanjing Meng}, {and}
  \bibinfo{person}{Ping Wang}.} \bibinfo{year}{2017}\natexlab{}.
\newblock \showarticletitle{Logdc: Problem diagnosis for declartively-deployed
  cloud applications with log}. In \bibinfo{booktitle}{\emph{2017 IEEE 14th
  International Conference on e-Business Engineering (ICEBE)}}. IEEE,
  \bibinfo{pages}{282--287}.
\newblock


\bibitem[Xu et~al\mbox{.}(2021)]%
        {xu2021care}
\bibfield{author}{\bibinfo{person}{Yong Xu}, \bibinfo{person}{Xu Zhang},
  \bibinfo{person}{Chuan Luo}, \bibinfo{person}{Si Qin}, \bibinfo{person}{Rohit
  Pandey}, \bibinfo{person}{Chao Du}, \bibinfo{person}{Qingwei Lin},
  \bibinfo{person}{Yingnong Dang}, {and} \bibinfo{person}{Andrew Zhou}.}
  \bibinfo{year}{2021}\natexlab{}.
\newblock \showarticletitle{CARE: Infusing causal aware thinking to root cause
  analysis in cloud system}. In \bibinfo{booktitle}{\emph{Proceedings of the
  1st Workshop on High Availability and Observability of Cloud Systems}}.
  \bibinfo{pages}{1--3}.
\newblock


\bibitem[{Xueqiu}(2023)]%
        {Vipshop2023}
\bibfield{author}{\bibinfo{person}{{Xueqiu}}.} \bibinfo{year}{2023}\natexlab{}.
\newblock \bibinfo{booktitle}{\emph{{Vipshop Nansha Data Center Failure}}}.
\newblock


\bibitem[Yao et~al\mbox{.}(2022)]%
        {yao2022react}
\bibfield{author}{\bibinfo{person}{Shunyu Yao}, \bibinfo{person}{Jeffrey Zhao},
  \bibinfo{person}{Dian Yu}, \bibinfo{person}{Nan Du}, \bibinfo{person}{Izhak
  Shafran}, \bibinfo{person}{Karthik Narasimhan}, {and} \bibinfo{person}{Yuan
  Cao}.} \bibinfo{year}{2022}\natexlab{}.
\newblock \showarticletitle{React: Synergizing reasoning and acting in language
  models}.
\newblock \bibinfo{journal}{\emph{arXiv preprint arXiv:2210.03629}}
  (\bibinfo{year}{2022}).
\newblock


\bibitem[Yu et~al\mbox{.}(2021)]%
        {2021MicroRank}
\bibfield{author}{\bibinfo{person}{Guangba Yu}, \bibinfo{person}{Pengfei Chen},
  \bibinfo{person}{Hongyang Chen}, \bibinfo{person}{Zijie Guan},
  \bibinfo{person}{Zicheng Huang}, \bibinfo{person}{Linxiao Jing},
  \bibinfo{person}{Tianjun Weng}, \bibinfo{person}{Xinmeng Sun}, {and}
  \bibinfo{person}{Xiaoyun Li}.} \bibinfo{year}{2021}\natexlab{}.
\newblock \showarticletitle{MicroRank: End-to-End Latency Issue Localization
  with Extended Spectrum Analysis in Microservice Environments}. In
  \bibinfo{booktitle}{\emph{WWW '21: The Web Conference 2021}}.
\newblock


\bibitem[Yu et~al\mbox{.}(2023a)]%
        {yu2023nezha}
\bibfield{author}{\bibinfo{person}{Guangba Yu}, \bibinfo{person}{Pengfei Chen},
  \bibinfo{person}{Yufeng Li}, \bibinfo{person}{Hongyang Chen},
  \bibinfo{person}{Xiaoyun Li}, {and} \bibinfo{person}{Zibin Zheng}.}
  \bibinfo{year}{2023}\natexlab{a}.
\newblock \showarticletitle{Nezha: Interpretable Fine-Grained Root Causes
  Analysis for Microservices on Multi-modal Observability Data}. In
  \bibinfo{booktitle}{\emph{Proceedings of the 31st ACM Joint European Software
  Engineering Conference and Symposium on the Foundations of Software
  Engineering}}. \bibinfo{pages}{553--565}.
\newblock


\bibitem[Yu et~al\mbox{.}(2023b)]%
        {yu2023tracerank}
\bibfield{author}{\bibinfo{person}{Guangba Yu}, \bibinfo{person}{Zicheng
  Huang}, {and} \bibinfo{person}{Pengfei Chen}.}
  \bibinfo{year}{2023}\natexlab{b}.
\newblock \showarticletitle{TraceRank: Abnormal service localization with
  dis-aggregated end-to-end tracing data in cloud native systems}.
\newblock \bibinfo{journal}{\emph{Journal of Software: Evolution and Process}}
  \bibinfo{volume}{35}, \bibinfo{number}{10} (\bibinfo{year}{2023}),
  \bibinfo{pages}{e2413}.
\newblock


\bibitem[Yu et~al\mbox{.}(2023c)]%
        {yu2023cmdiagnostor}
\bibfield{author}{\bibinfo{person}{Qingyang Yu}, \bibinfo{person}{Changhua
  Pei}, \bibinfo{person}{Bowen Hao}, \bibinfo{person}{Mingjie Li},
  \bibinfo{person}{Zeyan Li}, \bibinfo{person}{Shenglin Zhang},
  \bibinfo{person}{Xianglin Lu}, \bibinfo{person}{Rui Wang},
  \bibinfo{person}{Jiaqi Li}, \bibinfo{person}{Zhenyu Wu}, {et~al\mbox{.}}}
  \bibinfo{year}{2023}\natexlab{c}.
\newblock \showarticletitle{CMDiagnostor: An Ambiguity-Aware Root Cause
  Localization Approach Based on Call Metric Data}. In
  \bibinfo{booktitle}{\emph{Proceedings of the ACM Web Conference 2023}}.
  \bibinfo{pages}{2937--2947}.
\newblock


\bibitem[Yuan et~al\mbox{.}(2010)]%
        {yuan2010sherlog}
\bibfield{author}{\bibinfo{person}{Ding Yuan}, \bibinfo{person}{Haohui Mai},
  \bibinfo{person}{Weiwei Xiong}, \bibinfo{person}{Lin Tan},
  \bibinfo{person}{Yuanyuan Zhou}, {and} \bibinfo{person}{Shankar Pasupathy}.}
  \bibinfo{year}{2010}\natexlab{}.
\newblock \showarticletitle{Sherlog: error diagnosis by connecting clues from
  run-time logs}. In \bibinfo{booktitle}{\emph{Proceedings of the fifteenth
  International Conference on Architectural support for programming languages
  and operating systems}}. \bibinfo{pages}{143--154}.
\newblock


\bibitem[Zawawy et~al\mbox{.}(2010)]%
        {Zawawy2010log}
\bibfield{author}{\bibinfo{person}{Hamzeh Zawawy}, \bibinfo{person}{Kostas
  Kontogiannis}, {and} \bibinfo{person}{John Mylopoulos}.}
  \bibinfo{year}{2010}\natexlab{}.
\newblock \showarticletitle{Log filtering and interpretation for root cause
  analysis}. In \bibinfo{booktitle}{\emph{2010 IEEE International Conference on
  Software Maintenance}}. IEEE, \bibinfo{pages}{1--5}.
\newblock


\bibitem[Zhang et~al\mbox{.}(2023b)]%
        {zhang2023pace}
\bibfield{author}{\bibinfo{person}{Dylan Zhang}, \bibinfo{person}{Xuchao
  Zhang}, \bibinfo{person}{Chetan Bansal}, \bibinfo{person}{Pedro Las-Casas},
  \bibinfo{person}{Rodrigo Fonseca}, {and} \bibinfo{person}{Saravan Rajmohan}.}
  \bibinfo{year}{2023}\natexlab{b}.
\newblock \showarticletitle{PACE: Prompting and Augmentation for Calibrated
  Confidence Estimation with GPT-4 in Cloud Incident Root Cause Analysis}.
\newblock \bibinfo{journal}{\emph{arXiv preprint arXiv:2309.05833}}
  (\bibinfo{year}{2023}).
\newblock


\bibitem[Zhang et~al\mbox{.}(2023a)]%
        {zhang2023robust}
\bibfield{author}{\bibinfo{person}{Shenglin Zhang}, \bibinfo{person}{Pengxiang
  Jin}, \bibinfo{person}{Zihan Lin}, \bibinfo{person}{Yongqian Sun},
  \bibinfo{person}{Bicheng Zhang}, \bibinfo{person}{Sibo Xia},
  \bibinfo{person}{Zhengdan Li}, \bibinfo{person}{Zhenyu Zhong},
  \bibinfo{person}{Minghua Ma}, \bibinfo{person}{Wa Jin}, {et~al\mbox{.}}}
  \bibinfo{year}{2023}\natexlab{a}.
\newblock \showarticletitle{Robust failure diagnosis of microservice system
  through multimodal data}.
\newblock \bibinfo{journal}{\emph{IEEE Transactions on Services Computing}}
  (\bibinfo{year}{2023}).
\newblock


\bibitem[Zhang et~al\mbox{.}(2021)]%
        {zhang2021cloudrca}
\bibfield{author}{\bibinfo{person}{Yingying Zhang}, \bibinfo{person}{Zhengxiong
  Guan}, \bibinfo{person}{Huajie Qian}, \bibinfo{person}{Leili Xu},
  \bibinfo{person}{Hengbo Liu}, \bibinfo{person}{Qingsong Wen},
  \bibinfo{person}{Liang Sun}, \bibinfo{person}{Junwei Jiang},
  \bibinfo{person}{Lunting Fan}, {and} \bibinfo{person}{Min Ke}.}
  \bibinfo{year}{2021}\natexlab{}.
\newblock \showarticletitle{CloudRCA: A root cause analysis framework for cloud
  computing platforms}. In \bibinfo{booktitle}{\emph{Proceedings of the 30th
  ACM International Conference on Information \& Knowledge Management}}.
  \bibinfo{pages}{4373--4382}.
\newblock


\bibitem[Zhang et~al\mbox{.}(2022)]%
        {zhang2022crisp}
\bibfield{author}{\bibinfo{person}{Zhizhou Zhang},
  \bibinfo{person}{Murali~Krishna Ramanathan}, \bibinfo{person}{Prithvi Raj},
  \bibinfo{person}{Abhishek Parwal}, \bibinfo{person}{Timothy Sherwood}, {and}
  \bibinfo{person}{Milind Chabbi}.} \bibinfo{year}{2022}\natexlab{}.
\newblock \showarticletitle{$\{$CRISP$\}$: Critical Path Analysis of
  $\{$Large-Scale$\}$ Microservice Architectures}. In
  \bibinfo{booktitle}{\emph{2022 USENIX Annual Technical Conference (USENIX ATC
  22)}}. \bibinfo{pages}{655--672}.
\newblock


\bibitem[Zhao et~al\mbox{.}(2020)]%
        {zhao2020understanding}
\bibfield{author}{\bibinfo{person}{Nengwen Zhao}, \bibinfo{person}{Junjie
  Chen}, \bibinfo{person}{Xiao Peng}, \bibinfo{person}{Honglin Wang},
  \bibinfo{person}{Xinya Wu}, \bibinfo{person}{Yuanzong Zhang},
  \bibinfo{person}{Zikai Chen}, \bibinfo{person}{Xiangzhong Zheng},
  \bibinfo{person}{Xiaohui Nie}, \bibinfo{person}{Gang Wang}, {et~al\mbox{.}}}
  \bibinfo{year}{2020}\natexlab{}.
\newblock \showarticletitle{Understanding and handling alert storm for online
  service systems}. In \bibinfo{booktitle}{\emph{Proceedings of the ACM/IEEE
  42nd International Conference on Software Engineering: Software Engineering
  in Practice}}. \bibinfo{pages}{162--171}.
\newblock


\bibitem[Zhao et~al\mbox{.}(2023)]%
        {zhao2023design}
\bibfield{author}{\bibinfo{person}{Yibing Zhao}, \bibinfo{person}{Yongkun
  Zheng}, \bibinfo{person}{Haoran Luo}, \bibinfo{person}{Dengrong Wei},
  \bibinfo{person}{Chun Liu}, {and} \bibinfo{person}{Kang Chen}.}
  \bibinfo{year}{2023}\natexlab{}.
\newblock \showarticletitle{Design and Implement of AIOps System Based on
  Knowledge Graph}. In \bibinfo{booktitle}{\emph{2023 5th International
  Conference on Electronics and Communication, Network and Computer Technology
  (ECNCT)}}. IEEE, \bibinfo{pages}{285--288}.
\newblock


\bibitem[Zheng et~al\mbox{.}(2024)]%
        {zheng2024multi}
\bibfield{author}{\bibinfo{person}{Lecheng Zheng}, \bibinfo{person}{Zhengzhang
  Chen}, \bibinfo{person}{Jingrui He}, {and} \bibinfo{person}{Haifeng Chen}.}
  \bibinfo{year}{2024}\natexlab{}.
\newblock \showarticletitle{Multi-modal Causal Structure Learning and Root
  Cause Analysis}.
\newblock \bibinfo{journal}{\emph{arXiv preprint arXiv:2402.02357}}
  (\bibinfo{year}{2024}).
\newblock


\bibitem[Zhou et~al\mbox{.}(2015)]%
        {zhou2015distance}
\bibfield{author}{\bibinfo{person}{Junzan Zhou}, \bibinfo{person}{Shanping Li},
  {et~al\mbox{.}}} \bibinfo{year}{2015}\natexlab{}.
\newblock \showarticletitle{Distance based root cause analysis and change
  impact analysis of performance regressions}.
\newblock \bibinfo{journal}{\emph{Mathematical Problems in Engineering}}
  \bibinfo{volume}{2015} (\bibinfo{year}{2015}).
\newblock


\bibitem[Zhou et~al\mbox{.}(2023)]%
        {zhou2023tracestream}
\bibfield{author}{\bibinfo{person}{Tong Zhou}, \bibinfo{person}{Chenxi Zhang},
  \bibinfo{person}{Xin Peng}, \bibinfo{person}{Zhenghui Yan},
  \bibinfo{person}{Pairui Li}, \bibinfo{person}{Jianming Liang},
  \bibinfo{person}{Haibing Zheng}, \bibinfo{person}{Wujie Zheng}, {and}
  \bibinfo{person}{Yuetang Deng}.} \bibinfo{year}{2023}\natexlab{}.
\newblock \showarticletitle{TraceStream: Anomalous Service Localization based
  on Trace Stream Clustering with Online Feedback}. In
  \bibinfo{booktitle}{\emph{2023 IEEE 34th International Symposium on Software
  Reliability Engineering (ISSRE)}}. IEEE, \bibinfo{pages}{601--611}.
\newblock


\bibitem[Zhou et~al\mbox{.}(2019)]%
        {zhou2019latent}
\bibfield{author}{\bibinfo{person}{Xiang Zhou}, \bibinfo{person}{Xin Peng},
  \bibinfo{person}{Tao Xie}, \bibinfo{person}{Jun Sun}, \bibinfo{person}{Chao
  Ji}, \bibinfo{person}{Dewei Liu}, \bibinfo{person}{Qilin Xiang}, {and}
  \bibinfo{person}{Chuan He}.} \bibinfo{year}{2019}\natexlab{}.
\newblock \showarticletitle{Latent error prediction and fault localization for
  microservice applications by learning from system trace logs}. In
  \bibinfo{booktitle}{\emph{Proceedings of the 2019 27th ACM Joint Meeting on
  European Software Engineering Conference and Symposium on the Foundations of
  Software Engineering}}. \bibinfo{pages}{683--694}.
\newblock


\end{thebibliography}

\appendix

\end{document}